\documentclass[lettersize,journal]{IEEEtran}
\usepackage{cite}
\usepackage{amsthm}

\newtheorem{theorem}{Theorem}

\usepackage{lineno,xcolor}
\usepackage{amsmath}
\usepackage{comment}
\usepackage{amsfonts}
\usepackage{algorithm}
\usepackage{algpseudocode} 
\usepackage{amssymb}
\usepackage{graphicx}
\usepackage{subfigure}
\usepackage{subcaption}

\newtheorem{remark}{Remark}
\newtheorem{definition}[theorem]{Definition}

\DeclareMathOperator*{\argmax}{arg\,max}
\DeclareMathOperator*{\argmin}{arg\,min}

\begin{document}
	\title{Action-List Reinforcement Learning Syndrome \\ Decoding for Binary Linear Block Codes}
	\author{Milad~Taghipour, Bane~Vasi\'{c},~\IEEEmembership{Fellow,~IEEE}\\
		\IEEEauthorblockA{Department of Electrical and Computer Engineering, the University of Arizona, Tucson, AZ}
		\\Email: miladt@arizona.edu, vasic@ece.arizona.edu
	}
	
	\maketitle
	\begin{abstract}
		This paper explores the application of reinforcement learning techniques to enhance the performance of decoding of linear block codes based on flipping bits and finding optimal decisions. 
		We 
		describe the methodology for mapping the iterative decoding process into Markov Decision Processes (MDPs) and propose different methods to reduce the number of states in the MDP. A truncated MDP is proposed to reduce the number of states in the MDP by learning a Hamming ball with a specified radius around codewords. We then propose a general 
		scheme for reinforcement learning based decoders 
		applicable to any class of codes to improve the performance of decoders. We call this scheme an \emph{action-list decoding}. We design an action-list decoder based on the Deep-Q network values that substantially enhance performance. We also get benefit of automorphism group of code to further improve the code performance. Additionally, we propose a feedback-based method to exploit and enhance the performance of existing high-performing decoders by applying reinforcement learning algorithms after the existing decoders. These approaches effectively reduces the complexity of the reinforcement learning block. Finally, we present experimental results for the Low-Density Parity Check (LDPC) codes over the Binary Symmetric Channel (BSC) to demonstrate the efficiency of the proposed methods.
		
	\end{abstract}	
	
	\begin{IEEEkeywords}
		Reinforcement Learning, Action-List Decoding, Beam Decoding, Automorphism Group, Iterative Decoders, Linear Block Codes, QC-LDPC Codes, Bit Flipping.
	\end{IEEEkeywords}
	
	\section{Introduction}
	\IEEEPARstart{E}{rror} correcting codes are essential for ensuring reliable communication over noisy channels. Advances in coding theory have produced highly efficient codes and decoders. However, Maximum Likelihood (ML) decoding methods, such as the Viterbi \cite{forney1450960} and BCJR \cite{BCJR1055186} algorithms, are NP-hard for binary linear block codes, making them computationally infeasible for long-length or low-rate codes. To overcome these limitations, various approximate decoding techniques have been developed that offer lower complexity while approaching ML performance. These include Ordered Statistics Decoding (OSD) \cite{fossorier_osd_412683}, Polar Successive Cancellation List (SCL) decoding \cite{vardy_polar_7055304}, and message-passing algorithms based on factor graphs \cite{Kschischang910572}. While these methods perform well for long, well-structured LDPC codes, their effectiveness declines for short to moderate block lengths and high-density parity-check (HDPC) codes. Additionally, decoding strategies for convolutional codes using list decoding have been shown to reduce average complexity while maintaining near-optimal performance, particularly when leveraging techniques such as code decomposition and dual list decoding \cite{wesel10619434, Liva10144770}. 


	Recent advances in artificial intelligence, particularly in natural language processing and image processing, have led to increased interest in designing effective decoders for finite-length codes to enhance their decoding performance. Designing high-performance, fast, and low-complexity decoders for error-correcting codes remains a significant challenge, particularly for short to medium-length codes across general classes of codes. To bridge the gap between existing decoders and maximum likelihood (ML) decoding, one approach involves parameterized message-passing based decoders \cite{NachmaniDL, gruber2017deep, 8437530, 8006751, XinRNNFAID9057584}, which retain the advantages of BP while improving convergence speed and error performance. The authors in \cite{10418200} presented theoretical results associated with the generalization gap of the Neural BP decoder. Another approach is to view the problem of decoding as a decision-making process, using reinforcement learning (RL) to learn optimal strategies. Some works formulate decoding as a Markov decision process (MDP), where RL guides sequential bit-flipping decisions \cite{carpi2019reinforcement, rl_decoding}. In \cite{schedulingHabib}, the authors explore the scheduling problem in the belief propagation algorithm, framing it as a MDP and optimizing check-node scheduling policies to improve sequential decoding performance compared to traditional flooding scheduling methods. Throughout this paper, we propose a general and systematic method to reduce the state space size of the MDP. First, by focusing on the most probable error patterns and utilizing the concept of Hamming balls around codewords. In the coding theory literature, this concept is analogous to bounded distance decoding of linear block codes \cite{lin1989error}. 
	We observe that the Deep Q-learning algorithm demonstrates the ability to generalize and learn the optimal policy for certain states it has never encountered during training as demonstrated in Section \ref{sec:num_result}. We also propose a novel method to enhance the inference performance of a learned Deep Q-Network for general classes of binary linear codes. Due to the approximation of Q-values in the Deep Q-Network, the learned policy may exhibit oscillatory behavior and fail to converge. To address this issue, our proposed method employs a list of candidate solutions to guide the decoder more effectively and reduce the likelihood of convergence failures. Next, we propose a novel feedback decoding strategy designed to further enhance the performance of well-established decoders in the literature. This approach leverages an RL block to address error patterns that the chosen decoder fails to correct, thereby effectively expanding its correctable region. In general, this method can be employed to improve state-of-the-art iterative decoders and strengthen their error correction capabilities. Then, we use the automorphism group of the code to further improve the code performance. 
	At the first sight, it may appear counterintuitive that for the BSC permuting bit positions within a codeword would lead to any benefit since for the BSC all initial bit likelihood have the same magnitudes. However, we demonstrate that the sequential nature of reinforcement learning introduces anisotropy--position dependent bit-flipping rules--making the effect of an error pattern dependent on its position within the received codeword block. The remainder of this paper is organized as follows. Section \ref{sec:ch_coding} provides the necessary background on channel coding. Section \ref{sec:MDP} introduces the MDP framework and its application to decoding problems. In Section \ref{sec:bitflip_strategy}, we outline the RL-based bit-flipping strategy, followed by the proposed truncated MDP in Section \ref{sec:TruncMDP}. In Section \ref{sec:reward-analysis}, we investigate the effect of the discount factor on the reward function. Then, we introduce the action-list decoding technique to improve the decoder performance in Section \ref{sec:beam-decoder}. In Section \ref{sec:Feedback_dec}, the RL feedback decoding technique presented to improve the decoder performance. The automorphism group structure of the QC-LDPC code is presented in Section \ref{sec:automorphism_QC-LDPC}. Numerical results and performance comparisons are presented in Section \ref{sec:num_result}. The paper concludes with a summary of findings in Section \ref{sec:Conclusions}.

	\section{Preliminaries -- Channel Coding Background}
	\label{sec:ch_coding}
	Let us assume $\mathcal{C}$ to be a binary linear code with parameters $(n, k, d_{\textrm{min}})$, where $n$ is the code length, $k$ is the code dimension, and $d_{\textrm{min}}$ is the minimum distance of the code. We define a binary linear code $\mathcal{C} = \{\mathbf{c} \in \mathbb{F}_2^n \mid \mathbf{c}H^{T}=0 \}$, where $H$ is the parity-check matrix, which is an $m \times n$ binary matrix where $m/n \geq 1-R$, $R=k/n$ is the code rate. We also define $\mathbf{h}_{c_i}$ as the $i$-th column of the parity check matrix $H$. The error correction capability of a code can be defined on the basis of its minimum distance as $t = \lfloor {(d_{\textrm{min}}-1)}/{2} \rfloor$.
	A codeword $\mathbf{c}$ is transmitted through the Binary Symmetric Channel (BSC) as $\mathbf{y} = (1 - 2 \mathbf{c}) + \mathbf{z}$, where $\mathbf{y}$ is the received row vector of length $n$, and $\mathbf{z}$ is a row vector of length $n$ generated by i.i.d. Bernoulli random variable with parameter $\rho$ which is the cross-over probability of the channel.
	
	\subsection{Bounded Distance Decoding}
	Bounded distance decoding (BDD), studied in the error correction literature, aims to correct errors of weight up to $w$. The \emph{frame error rate} (FER) of BDD over the binary symmetric channel (BSC) is given by
	\begin{equation}
		P_{\textrm{FER}} (w) = 1 - \sum_{i=0}^{w} \binom{n}{i} \rho^i (1-\rho)^{n-i}.
		\label{eq::BDD_error}
	\end{equation}
	We utilize the BDD frame error probability, as expressed in Equation \ref{eq::BDD_error}, as a baseline to evaluate the performance improvements achieved by our proposed methods in the Numerical Results section.
	
	\subsection{Error Floor Estimation}
	In this Section, we discuss the estimation of the Error-Floor of LDPC codes. The probability of frame error can be expressed in terms of the errors that the decoder fails to correct as
	\begin{equation}
		P_{FER}(\rho) = \sum_{i=c}^n \mid E_i \mid \rho^i (1-\rho)^{n-i}
	\end{equation}
	where $|E_i|$ denotes the number of weight-$i$ error patterns that the decoder is unable to correct, and $c$ denotes the minimum weight of an uncorrectable error pattern. For small values of $\alpha$, the probability of error can be approximated by considering only the contribution of the lowest-weight uncorrectable error patterns. This allows for estimating the decoder’s performance on a semi-log scale in the error floor region as
	\begin{equation}
		\log \lim_{\rho \to 0} P_{FER}(\rho) = \log |E_i| + c \log \rho.
	\end{equation}
	We later use this approximation to demonstrate how our proposed feedback model improves decoder performance in terms of Frame Error Rate (FER).

	\section{Preliminaries --Markov Decision Process}
	\label{sec:MDP}
	Markov Decision Processes (MDPs) provides a mathematical framework for modeling decision-making problems where outcomes are random or deterministic under the control of a decision maker.
	An MDP can be defined based on $(S,A,P,R,\gamma)$, where $S$ is the set of finite states representing all situations of the environment, $A$ is the set of finite actions that the decision maker, or agent, can take, $P_{s s^\prime}^{a} = P(s^\prime \mid s, a ) \triangleq P(S_{t+1}=s^{\prime} \mid S_t=s, A_t=a) $ is the probability transition function $P:S\times A \times S \rightarrow [0,1]$ which represents the probability of going to state $s^\prime$ from state $s$ after taking action $a$, $R(s,a)$ is the reward function $R: S\times A \rightarrow \mathbb{R}$ which represents the immediate reward received after taking action $a$ in state $s$, and $\gamma \in [0,1]$ is the discount factor which denotes the importance of future rewards. A deterministic policy $\pi$ is a mapping $\pi: S \rightarrow A$, where $\pi(s)$ denotes the action to take in state $s$. The goal in an MDP is to find an optimal policy $\pi^{\star}$ that maximizes the expected cumulative reward $\pi^{\star} = \argmax_{\pi} \mathbb{E}_{\pi} [ \sum_{t=0}^{\infty} \gamma^t R(s_t, \pi(s_t)) \mid s_0 = s ]$, where $\mathbb{E}_{\pi}$ is the expectation by taking actions under policy $\pi$.
	
	
	\subsection{Q-learning}
	\label{subsec:Q-learning}
	Q-learning is a model-free reinforcement learning algorithm that is used to solve MDPs. It has the ability to learn the optimal policy $\pi^{\star}$ directly from interactions with the environment without requiring the model of environment. The action-value function $Q:S \times A \rightarrow \mathbb{R}$ is defined to be the expected cumulative rewards starting from state $s$, taking action $a$, and using policy $\pi$ to select actions afterwards
	\begin{equation}
		Q_{\pi} (s,a) = \mathbb{E}_{\pi} [ \sum_{t=0}^{\infty} \gamma^t R(s_t, \pi(s_t)) \mid S_0 = s, A_0 = a ].
	\end{equation}
	Q-learning algorithm aims to learn the optimal action-value function
	\begin{equation}
		Q^{\star} (s,a) = \max_{\pi} Q_{\pi} (s,a)
	\end{equation}
	that find the expected reward of taking action $a$ in state $s$ and following the optimal policy afterwards. The Q-learning algorithm iteratively updates the Q-values using the following rule
	\begin{equation}
		Q(s,a) \leftarrow Q(s,a) + \alpha [R(s, a) + \gamma \max_{a^{\prime}} Q(s^{\prime}, a^{\prime})  - Q(s,a)]
	\end{equation}
	updating the Q-values based on the immediate rewards received and the estimated value of subsequent maximum action-value estimate where $\alpha \in (0,1]$ is the step size for updating the Q values.
	The Q-learning algorithm is shown in Algorithm \ref{alg:Q_algo} \cite{sutton2018reinforcement}, where $s_E$ is the terminal state. It is worth mentioning that the complexity of Q-learning algorithm is $O(|S| \times |A|^2 )$, due to the necessity of updating each state-action pair and calculating the maximum Q-value over all possible actions in the subsequent state during each iteration.
	\begin{algorithm}
		\caption{Q-learning Algorithm}
		\label{alg:Q_algo}
		\textbf{Input:} Initialize $Q(s, a)$ arbitrarily (e.g., randomly), except $Q(s_{\text{terminal}}, a) = 0$. Set the number of episodes $N_{\text{episodes}}$. \\
		\textbf{Output:} Optimal action-value function $Q(s, a)$ for all $s \in S$, $a \in A$.
		\begin{algorithmic}[1]
			\For{$i \gets 1$ to $N_{\text{episodes}}$}
			\State Initialize state $s$
			\While{$s$ is not terminal}
			\State Choose action $a$ from state $s$ using an exploration strategy (e.g., $\epsilon$-greedy)
			\State Take action $a$, observe reward $R(s, a)$ and next state $s'$
			\State Update $Q$-value:
			\[
			Q(s, a) \leftarrow Q(s, a) + \alpha \left[ R(s, a) + \gamma \max_{a'} Q(s', a') - Q(s, a) \right]
			\]
			\State $s \leftarrow s'$
			\EndWhile
			\EndFor
		\end{algorithmic}
	\end{algorithm}
	The selection of actions in the Q-learning algorithm is an important step that needs to be balanced between exploration, experiencing new actions, and exploitation, choosing the best known action. The $\epsilon$-greedy strategy is commonly employed for this purpose. 
	It selects actions based on a probability distribution defined by the parameter $\epsilon$, as follows
	\begin{equation}
		a_t = 
		\begin{cases}
			\textrm{random action uniformly from } A, & \textrm{w.p. } \epsilon \\
			\argmax_{a \in A} Q(s_t, a) , & \textrm{w.p. } 1-\epsilon
		\end{cases}
	\end{equation}
	A simple and effective way to reduce exploration in favor of exploitation as the learning process progresses is the use of linear decay in $\epsilon$-greedy algorithm. By controlling the decay rate, the agent can effectively balance between discovering new actions and leveraging learned knowledge to maximize rewards. The linear decaying $\epsilon (t)$ at the episode $t$ is given by
	\begin{equation}
		\epsilon(t) = \max(\epsilon_{min}, \epsilon_{max} - \dfrac{\epsilon_{max}-\epsilon_{min}}{N} \cdot t)
	\end{equation}
	where $\epsilon_{max}$/$\epsilon_{min}$ is the maximum/minimum value of $\epsilon$ and $N$ is the total number of episodes.
	
	\subsection{Deep Q-learning Algorithm}
	\label{subsec:DQN}
	Deep Q-learning is an extension of Q-learning algorithm that uses deep neural networks to approximate the Q-values \cite{sutton2018reinforcement}. We employ a neural network to estimate a low-complexity approximation of the Q-table, $Q(s,a)$. We assume $Q_{\theta}(s,a )$ represents the estimation of the Q-function with learnable parameters $\theta$. We store transition tuples $(s,a,r,s^{\prime})$ obtained from simulating the MDP in a set $\mathcal{B}$ called replay memory. For stable training, we use two neural networks a primary Q-network $Q_{\theta}^{p} (s,a)$ for selecting actions and a target Q-network $Q_{\theta}^{t} (s, a)$ to provide more stable Q-value updates. The agent balances exploration and exploitation using an $\epsilon$-greedy policy with linear decay. Then, we update the parameter $\theta$ based on reducing the loss function
	\begin{equation}
		L(\theta) = \mathbb{E}_{(s,a,r,s^{\prime}) \in \mathcal{B}} [r + \gamma \max_{a^{\prime} \in A} Q_{\theta^{-}}^{t} (s^{\prime}, a^{\prime}) - Q_{\theta}^{p} (s,a)]^2
		\label{eq:loss_deep}
	\end{equation}
	where gradient descent based methods are employed to update the parameters $\theta$ to minimize the loss function defined in Equation (\ref{eq:loss_deep}). Periodically, after some iterations the target network is updated to match the primary network $\theta^{-} \leftarrow \theta$, improving the stability of training. The goal is to maximize cumulative rewards by learning an optimal policy over time. The full algorithm is shown in Algorithm \ref{alg:deep_q_learning}.
	
	\begin{algorithm}
		\caption{Deep Q-learning Algorithm}
		\label{alg:deep_q_learning}
		
		\textbf{Input}: Initialize replay memory $\mathcal{B}$, initialize Q-networks $Q^{p}_{\theta}(s,a)$ and $Q^{t}_{\theta}(s,a)$ with random weights $\theta$ and $\theta^-$, and set \texttt{epo} as the number of episodes. \\
		\textbf{Output}: Primary Q-network $Q^{p}_{\theta}(s,a)$
		
		\begin{algorithmic}[1]
			\For{$i = 1, 2, \dots, \texttt{epo}$}
			\State Sample a random minibatch $(s, a, r, s')$ of transitions from $\mathcal{B}$.
			\State Set $y \leftarrow \begin{cases} 
				r, & \text{if } s' \in S_E \\
				r + \gamma \max\limits_{a'} Q^{t}_{\theta^-}(s', a'), & \text{otherwise}
			\end{cases}$
			\State Perform a gradient descent step on $\left(y - Q^{p}_{\theta}(s,a)\right)^2$ with respect to $\theta$.
			\If {fixed number of steps are reached}
			\State Update $Q^{t} \leftarrow Q^{p}$
			\EndIf
			\EndFor
		\end{algorithmic}
	\end{algorithm}

	\section{Bit-flipping Strategy}
	\label{sec:bitflip_strategy}
	Assuming transmission over independent memoryless channels, hard-decision maximum likelihood decoding of binary linear block codes can be interpreted as finding the minimum number of bit flips in the received word required to obtain the all-zero syndrome.
	
	\subsection{State and Action Spaces}
	\label{subsec:stateActionSpaces}
	Given that the code dimension is $n$, each of these $n$ bits can be erroneous. In other words, the action space $A$ should be designed to allow the flipping of each bit at any time step. Therefore, we can state $A=\{1,2,\cdots, n \}$. 
	The decoding process can be terminated based on the value of the syndrome vector, which indicates the positions of unsatisfied checks in the code. Consequently, the state space can be defined by all possible binary syndromes of length $m$. The initial state $\mathbf{s}_0 = \mathbf{y} H^T $ represents the syndrome of the received vector. Next, we need to define the transition probability function $P(\mathbf{s}^{\prime} \mid \mathbf{s}, a)$ for an MDP. By taking action $a$ from state $\mathbf{s}$, we transition to state $\mathbf{s}^{\prime} = \mathbf{s} + \mathbf{h}_{c_a}$ with probability $1$, indicating that it is a deterministic MDP.
	Terminal state corresponds to the all-zero syndrome, indicating that a codeword is obtained and that the decoding process should be terminated. We also impose a limit of $L$ steps, which is the maximum length of each episode. If an episode is not terminated within $L$ steps, a new episode begins.

	\subsection{Reward Strategy}
	Choosing a reward strategy is crucial in Reinforcement Learning algorithms because it significantly impacts the learning efficiency and the resulting policy. 
	Since the goal of the decoding problem is to find a codeword, we define a positive reward of $1$ if an agent reaches the terminal state.
	It is also necessary to penalize the action of flipping a bit at each step in order to encourage the decoder to find a solution that requires the fewest possible bit flips. This helps identifying the set of least reliable bits to find a codeword, thereby minimizing the probability of word error rate.
	Thus, we define a reward function as in \cite{carpi2019reinforcement} for the BSC
	\begin{equation}
		R(s,a,s^{\prime}) = 
		\begin{cases}
			-\frac{1}{L} +1, \quad & \textrm{if } s^{\prime} = 0  \\
			-\frac{1}{L},   & \textrm{otherwise.}
		\end{cases}
	\end{equation}
	
	\section{Truncated MDP}
	\label{sec:TruncMDP}
	The proposed Truncated Markov Decision Process (TMDP) faces challenges related to the number of states, which corresponds to the number of syndromes of the code, given by $2^{n-k}$. This number grows exponentially, making Q-Learning methods infeasible for long and low-rate codes. In this section, we propose a reduction in the state space size of the MDP model aimed at handling low-rate codes to learn their optimal policies. We use state reduction in the proposed MDP to make the problem more tractable. We define a Hamming ball $\mathcal{B}(c, w)$ with a radius $w$ for a binary linear code of length $n$ to cover all received words within a maximum hamming distance of $w$ or less around a codeword $c \in \mathcal{C}$ defined as
	\begin{equation}
		\mathcal{B}(c, w) = \{ \mathbf{x} \in \mathbb{F}_2^n \mid d_H (\mathbf{x}, c) \leq w \}
	\end{equation}
	where $d_H(\mathbf{x}, c)$ represents the hamming distance between the codeword $c$ and the received word $\mathbf{x}$. We define a hamming ball $\mathcal{B}(w)$ with radius $w$ of a code $\mathcal{C}$ to be the union of hamming balls centered at each codeword $c \in \mathcal{C}$, i.e., $\mathcal{B}(w) = \cup_{\forall c\in \mathcal{C}} \mathcal{B}(c, w)$. The hamming ball $B(w)$ includes all received words within a maximum hamming distance of $w$ from any codeword $c \in \mathcal{C}$. $S(w)$ denotes the set of syndromes associated with $\mathcal{B}(w)$, where $S(w) = \{\mathbf{y} H^T \mid \mathbf{y} \in \mathcal{B}(w)\}$.
	
	To evaluate the performance of a coding and decoding scheme for linear codes over an output symmetric channel, it is sufficient to assume the transmission of the all-zero codeword, provided that the decoding algorithm satisfies certain conditions, including channel symmetry and decoder symmetry (Lemma $1$ in \cite{richardson2001capacity}). The iterative decoding algorithms considered throughout this paper satisfy these symmetry conditions, making both the learning of decision regions and the probability of error independent of the transmitted codeword. Therefore, we can assume the transmission of the all-zero codeword. 
	Binary linear block codes may have complex decision regions that make finding the optimal policy difficult. It is known that the error correction capability of the code is a good estimate for its probability of error in a high SNR regime. Therefore, we consider only the states corresponding to syndromes related to erroneous patterns up to weight $w$. With this modification, we dramatically decrease the number of states to
	\begin{equation}
		\mid S \mid = \sum_{i=0}^{w} \binom{n}{i}
	\end{equation}
	which is significantly smaller than the total number of syndromes. The action space can be modified for the states on the boundary (weight-$w$ error patterns) to be a specific set related to their syndrome. Alternatively, the action space can remain as before, with a new reward of $-1$ assigned for actions that transition to states outside our favorable state space. The proposed reward function can be defined as follows
	\begin{equation}
		R(s,a,s^{\prime}) = 
		\begin{cases}
			-\frac{1}{L} +1, \quad & \textrm{if } s^{\prime} = \mathbf{0}  \\
			-\frac{1}{L},   & \textrm{is } s^{\prime} \in S(w)\backslash \mathbf{0} \\
			-\frac{1}{L} -1, & \textrm{if } s^{\prime} \in \overline{S(w)}
		\end{cases}
	\end{equation}
	where $\overline{S(w)}$ denotes the complement of the set $S(w)$. 
	In the proposed method, we aim to correct up to weight-$w$ errors, which is closely related to the concept of bounded distance decoding (BDD) in the error correction literature, given by Equation \ref{eq::BDD_error}, as a baseline to evaluate the accuracy of the found optimal policy. In the proposed truncated Markov Decision Process (MDP), we intuitively define a bounded-distance decoding version of the MDP, which, at best, achieves the performance corresponding to the bounded-distance error probability. In the subsequent section, we analyze the number of optimal policies that the it can be found.

	\subsection{Number of Optimal Policies}
	The Q-learning algorithm aims to determine the optimal policy $\pi^{\star}$. In this section, we analyze the existence and combinatorial properties of optimal policies that Q-learning ultimately tries to identify for decoding binary linear block codes. Under certain conditions, the Q-learning algorithm converges to the optimal policy, as stated in \cite{sutton2018reinforcement}. Given the convergence of the Q-learning algorithm, we can formulate the following theorem regarding the number of optimal policies that the algorithm can identify for decoding binary linear codes. In the subsequent section, we analyze the number of optimal policies that can be found.
	
	
	\begin{theorem}
		The number of distinct optimal policies, or proposed decoders, for the Binary Symmetric channel is at least
		\begin{equation}
			\prod_{i=1}^{t} i^{\binom{n}{i}} 
			\label{eq:nopt_pol}
		\end{equation}
		where $n$ denotes the codelength and $t$ represents the error-correcting capability of the code.
	\end{theorem}
	
	\begin{proof}
		The proof is provided in the Appendix \ref{Appendix:proof1}.
	\end{proof}
	
	\section{Finite Horizon MDP reward function}
	\label{sec:reward-analysis}
\begin{figure}
		\centering\includegraphics[width=0.8\columnwidth]{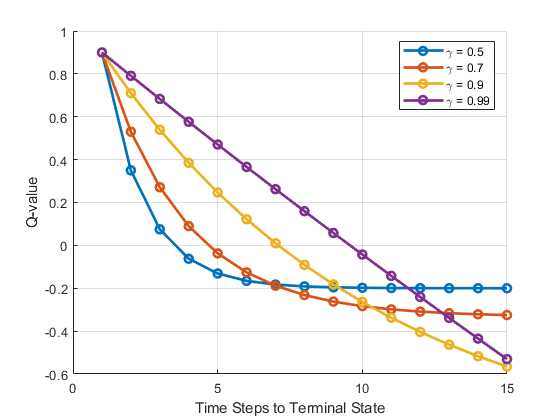}
		\caption{Q-values as a Function of Time Steps to Terminal State.}
		\label{fig:q_values}
	\end{figure}

	In reinforcement learning, the objective is to ensure that all states—including those that are far from terminal states—are learned efficiently. However, learning distant states is challenging due to low Q-values, vanishing gradients, and suboptimal exploration. For binary linear codes, the goal is to learn and correct error patterns up to the error-correcting capability of the code, i.e. $t$, and further. The expected return of a state-action pair, under a deterministic policy where the agent reaches the terminal state in exactly $j$ steps, can be expressed as
	\begin{equation}
		Q(j) = \gamma^{j-1} r - \dfrac{1-\gamma^j}{1-\gamma} p
	\end{equation}
	where $r$ is the terminal reward, $p$ is the per-step penalty value, and $\gamma$ is the discount factor. In Figure \ref{fig:q_values}, we set $r=1$ and $p=-1/10$ to illustrate the effect of the discount factor $\gamma$. For small values of $\gamma$, Q-values drop rapidly, meaning the agent strongly prefer short term rewards. In this case, the Q-values associated with states that are far from the terminal state become indistinguishably small, making it difficult for a Deep Q-Network (DQN) to learn meaningful representations of those states. In contrast, states that are close to the terminal state exhibit more distinct Q-values, allowing the DQN to learn them more effectively. On the other hand, for larger values of $\gamma$, the Q-values of distant states decay more slowly, preserving distinguishability and enabling the DQN to learn these states more effectively. However, in such cases, the Q-values of the initial states may become less distinguishable from one another compared to those under smaller values of $\gamma$, which can hinder the learning of precise value estimates for these early states.
	In the subsequent sections, we present general methodologies that leverage this intuitive analysis to design robust decoders for binary linear block codes.

	\section{Action-List decoding based on Q-values}
	\label{sec:beam-decoder}
\begin{figure}
		\centering\includegraphics[width=0.9\columnwidth]{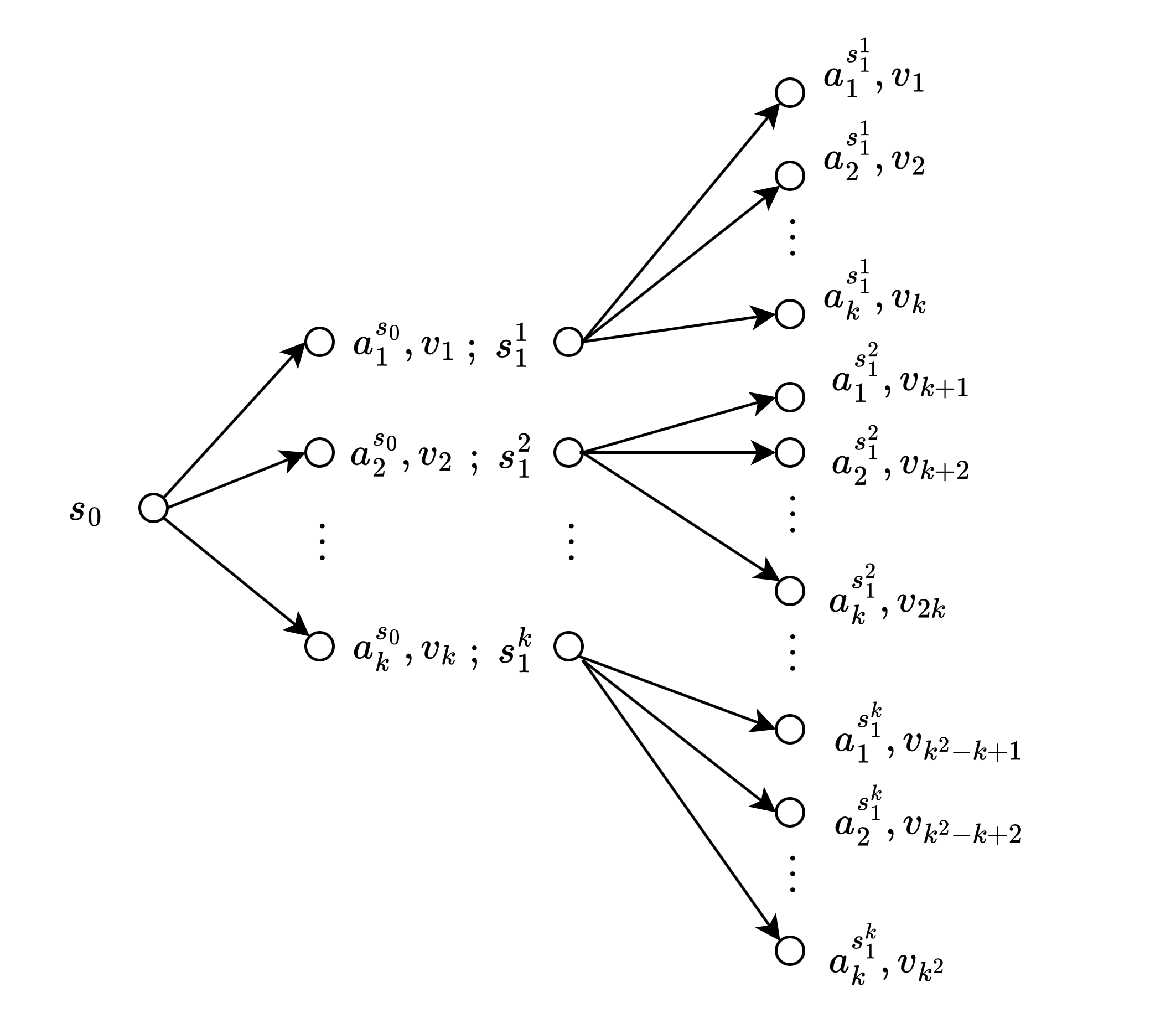}
		\caption{$2$-Depths expansion of Action-List Algorithm.}
		\label{fig:Beam_decoding_visualization}
	\end{figure}

	The proposed approach involves determining the optimal policy by leveraging Q-values through greedy action selection. However, due to the nature of the Deep Q-Network as a function approximator, there is no guarantee that it will learn the exact Q-values. Using the optimal policy, the decoder is then executed in an iterative manner to decode the received word. During the decoding phase, the decoder may become trapped in oscillatory behavior due to the limitations of the learned policy. The objective is to develop a robust decoder that can lower the probability of error. To improve decoder performance, we propose a new type of decoder in which, at each state, multiple actions are considered rather than selecting only the best possible action based on the Q-values. The proposed approach is conceptually similar to beam search optimization \cite{BeamSearchNEURIPS2022}. Its purpose is to mitigate the issue of the decoder becoming trapped in oscillatory behavior. We define a deterministic state transition function $T:\mathcal{S} \times \mathcal{A} \rightarrow \mathcal{S}$ such that $\mathbf{s}^{\prime} = T(\mathbf{s}, a) = \mathbf{s} + \mathbf{h}_{c_a}$, which maps the current state to the next state based on the selected action. Specifically, the next state is obtained by adding the current state to the $a$-th column of the parity-check matrix. We have an initial state $\mathbf{s}_0$. The Q-values for all actions in $\mathbf{s}_0$, denoted by $\{Q(\mathbf{s}_0, a) \mid a \in \mathcal{A} \}$, are computed using the Deep-Q Network. We then select the top $k$ actions with the highest Q-values. For each selected action $a$, we construct a candidate path $p^i$
	\begin{equation}
		p^i = \{\mathbf{s}_0^i, \mathbf{s}_1^i = T(\mathbf{s}_0, a)  \}
	\end{equation}
	where $i \in [k]$, and assign a score $v^i = Q(\mathbf{s}_0^i,a)$ to each candidate path.
	We assume that at depth $l$, we maintain a list of candidate paths denoted by $\mathcal{B}_l = \{ (p^i, v^i) \mid i \in [k]  \}$ where each $p^i$ is the sequence of states $(\mathbf{s}_0^i, \mathbf{s}_1^i, \cdots, \mathbf{s}_l^i)$ and $v^i$ is the Q-value of the last action that generated the state $s_l^i$. In the path expansion stage, for each candidate path $(p^i, v^i) \in \mathcal{B}_l$, we compute the Q-values at the current state $\mathbf{s}_l^i$. A path $p^i$ is extended only by actions $a$ satisfying the condition
	\begin{equation}
		Q(\mathbf{s}_l^i,a) > v^i
	\end{equation}
	which ensures that the Q-value of the new action is strictly greater than that of its parent. This criterion enforces the selection of improving paths, guiding the decoder toward convergence to the all-zero syndrome.
	For all actions that satisfy the condition, a new state is generated as $\mathbf{s}_{l+1} = T(\mathbf{s}_l, a)$, and an extended path is formed as $\hat{p}^{i} = p \cup\{ \mathbf{s}_{l+1} \} $, with an updated score $\hat{v}^i = Q(\mathbf{s}_l, a)$. In the path pruning stage, all extended paths generated from the candidates in $\mathcal{B}_l$ are collected into a new set. From this set, the top $k$ paths are selected based on their scores $\hat{v}$ to form the updated path collection $\mathcal{B}_{l+1}$, where $|\mathcal{B}_{l+1}| \leq k$, or fewer if not enough valid candidates are available. The search process terminates when at least one candidate path $p^i$ satisfies the all-zero syndrome, or when a maximum search depth $D_\textrm{max}$ is reached. The proposed method outputs the list of candidate paths that reach a valid codeword along with their associated scores which is presented in Algorithm \ref{alg:beam-search}. The candidate path with the highest estimated value is selected as the final output. 
	In Figure \ref{fig:Beam_decoding_visualization}, we illustrate the steps of the action-list algorithm up to depth $2$ for further clarification. We begin with the initial syndrome. Using the policy network, we sort actions according to their estimated Q-values. The top-$k$ actions are then selected as part of the expansion procedure, resulting in transitions to the next set of syndromes $\{s_1^1, s_1^2, \cdots, s_1^k\}$. From each of the states at depth-$1$, we again select the top-$k$ actions, yielding a total of $k^2$ expanded paths. As a pruning step, only the top-$k$ paths are used for further expansion. This process is repeated until either a valid codeword is found or the predefined number of steps is reached. In the action-list algorithm, the policy network is evaluated $k$ times at each depth, followed by $k$ sorting operations.
	In the next section, we aim to leverage the performance of existing well-designed decoders and refine their state space to improve their performance using Reinforcement Learning algorithms.

	\begin{algorithm}
		\caption{Q-Guided Action-List Decoding}
		\label{alg:beam-search}
		
		\begin{algorithmic}[1]
			\Require 
			Trained Q-network $Q(\cdot,\cdot)$,initial syndrome $\mathbf{s}_0$, list size $k$, maximum iterations $D_{\max}$, parity-check matrix $H$. 
			\Ensure 
			A candidate path $p^*$ that reaches a valid codeword.
			
			\State \textbf{Initialize} beam $\mathcal{B}_0 \gets \emptyset$.
			\State Compute Q-values for $\mathbf{s}_0$: \quad $q_0(a) \gets Q(\mathbf{s}_0,a),$  $\forall a \in \mathcal{A}$
			\State Let $A_0 \gets \text{Top}_k(\{q_0(a)\}_{a \in \mathcal{A}})$
			\For{each action $a \in A_0$}
			\State $\mathbf{s}_1 \gets T(\mathbf{s}_0,a)$
			\State \textbf{Add} candidate path $p = \{ \mathbf{s}_0, \mathbf{s}_1 \}$ with score $v = Q(\mathbf{s}_0,a)$ to $\mathcal{B}_0$
			\EndFor
			\State $i \gets 0$
			\While{$i < D_{\max}$ and $\mathcal{B}_i \neq \emptyset$}
			\State Initialize candidate set $\mathcal{T} \gets \emptyset$
			\For{each candidate $(p, v)$ in $\mathcal{B}_i$ with last state $\mathbf{s}$}
			\State Compute Q-values: \quad $q(\mathbf{s},a) \gets Q(\mathbf{s},a),$ $\forall a \in \mathcal{A}$
			\State Let $A \gets \text{Top}_k(\{q(\mathbf{s},a)\}_{a \in \mathcal{A}})$
			\For{each action $a \in A$}
			\If{$q(\mathbf{s}, a) > v$} \Comment{Only extend if improvement is observed.}
			\State $\mathbf{s}^{\prime} \gets T(\mathbf{s},a)$
			\State $p^{\prime} \gets p \cup \{\mathbf{s}^{\prime}\}$
			\State \textbf{Add} candidate $(p^{\prime}, q(\mathbf{s},a))$ to $\mathcal{T}$
			\EndIf
			\EndFor
			\EndFor
			\State Sort $\mathcal{T}$ in descending order by score.
			\State Let $\mathcal{B}_{i+1} \gets \text{Top}_k(\mathcal{T})$
			\For{each candidate $(p^{\prime}, v^{\prime})$ in $\mathcal{B}_{i+1}$}
			\If{$\text{Valid}(\mathbf{s}^{\prime})$ \textbf{is true}, where $\mathbf{s}^{\prime}$ is the last state in $p^{\prime}$}
			\State \Return $p^{\prime}$
			\EndIf
			\EndFor
			\State $i \gets i + 1$
			\EndWhile
			\State \Return Best candidate path from $\mathcal{B}_i$ (or \texttt{null} if none found)
		\end{algorithmic}
	\end{algorithm}

	\section{Feedback Decoder}
	\label{sec:Feedback_dec}
	
\begin{figure*}
		\centering\includegraphics[width=0.8\textwidth]{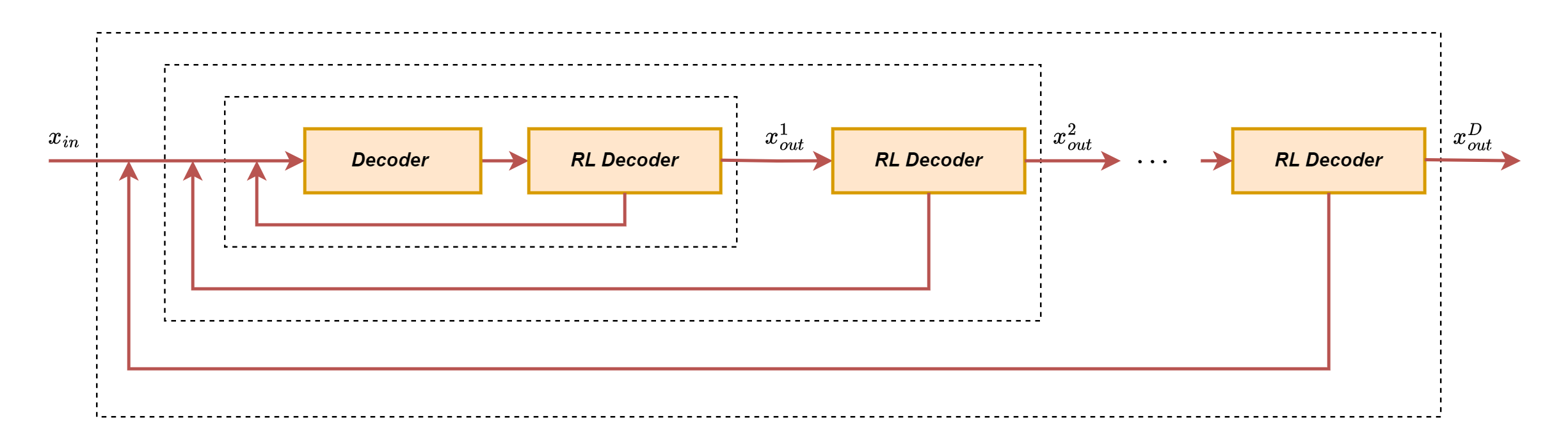}
		\caption{Feedback based Reinforcement Learning decoder.} 
		\label{fig:feedback_RL}
	\end{figure*}

\begin{figure}
		\centering\includegraphics[width=0.9\columnwidth]{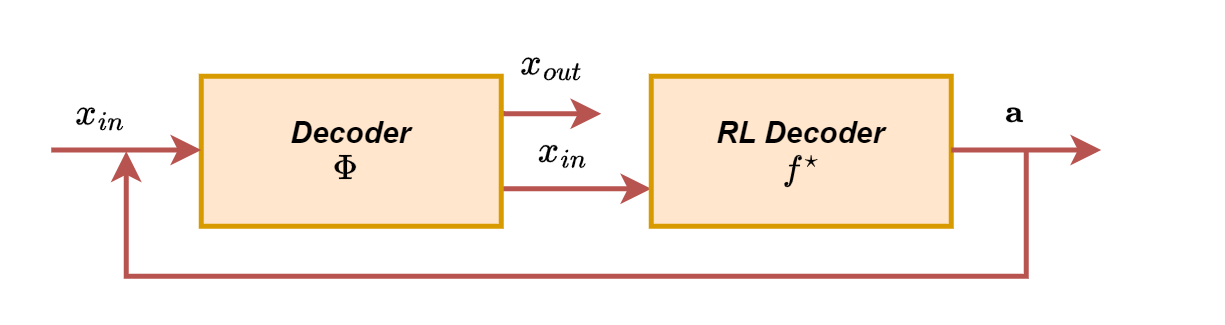}
		\caption{Feedback based Reinforcement Learning decoder.}
		\label{fig:BF_RL}
	\end{figure}

	As explained in the Section \ref{subsec:Q-learning}, the complexity of Q-learning algorithm is $O(|S| \times |A|^2)$, which is dependent on the number of states and the square of the number of actions. For the MDP defined in Section \ref{subsec:stateActionSpaces}, it becomes $O(n^2 2^{n-k})$. In this section, we aim to establish a general framework to reduce the number of states in the MDP. Since the approach described in \ref{sec:bitflip_strategy} becomes infeasible for long-length and low-rate codes, we propose a general feedback method to leverage the advantages of well-established codes and decoders from the literature. We incorporate a Reinforcement Learning block to correct errors that the chosen decoder cannot correct, thereby enhancing the decoder's performance through the RL-based bit flipping mechanism. In addition, this RL-based feedback decoder can be applied sequentially for $D$ times to further enhance performance as shown in Figure \ref{fig:feedback_RL}.
	Each RL block, as shown in Figure \ref{fig:BF_RL}, is designed to correct the failures of the employed decoder, thereby expanding the overall decoder's correctable region. We define $\mathcal{R}^{\Phi}_{c} $ to be the successful decoding region of a chosen decoder $\Phi$ 
	\begin{equation}
		\mathcal{R}^{\Phi}_{c} = \{ x \in \mathbb{F}^n_{2} | \Phi(x) = \argmin_{c^{\prime} \in \mathcal{C}} d(x, c^{\prime})  \}.
	\end{equation}
	We also define the miscorrectable region of the decoder $\Phi$ as the region in which the decoder assigns a codeword to the received word that is not the closest codeword.
	\begin{equation}
		\mathcal{R}^{\Phi}_{m} = \{ x \in \mathbb{F}^n_{2} | \Phi(x) \in \mathcal{C} , \Phi(x) \neq \argmin_{c^{\prime} \in \mathcal{C}} d(x, c^{\prime})  \}
	\end{equation}
	Then, we can define the set of failure-inducing error patterns as the complement of the set of correctable and miscorrectable error patterns
	\begin{equation}
		\mathcal{R}^{\Phi}_{f} = \{ x \in \mathbb{F}^n_{2} | \Phi(x) \notin \mathcal{R}^{\Phi}_{c} \cup \mathcal{R}^{\Phi}_{m}  \}.
	\end{equation}
	We denote by $S^{\Phi}_i$ the set of syndromes corresponding to the $i$ (correctable, failure, or miscorrectable) decoding region ${\mathcal{R}^{\Phi}_i }$ of the decoder $\Phi$, where $S^{\Phi}_i = \{\mathbf{y} H^T \mid \mathbf{y} \in {\mathcal{R}^{\Phi}_i} \}$ and $i \in \{c,f,m\}$. The states of new Markov Decision Process (MDP) are $S^{\Phi}_f \cup \{\mathbf{0}\}$, and the action space is defined as $ A = \{1,2,\cdots,n\}$. The reward function is defined as follows
	\begin{equation}
		R(s,a,s^{\prime}) = 
		\begin{cases}
			-\frac{1}{L} +1, \quad & \textrm{if } s^{\prime} \notin S^{\Phi}_f  \\
			-\frac{1}{L},   & \textrm{is } s^{\prime} \in S^{\Phi}_f.
		\end{cases}
		\label{eq:reward_feedbackDec}
	\end{equation}
	The RL block interacts with the existing decoder to expand its correctable region. The general decoding procedure during inference is outlined in Algorithm \ref{alg:FeedbackDec_algorithm}, where $\pi^{\star} (\mathbf{s})$ denotes the optimal policy. Given the convergence of the Q-learning algorithm, we formulate the following theorem to characterize the error-correction capability of the proposed feedback decoder under the convergence of the RL algorithm.
	
	\begin{algorithm}
		\caption{Decoder with Reinforcement Learning Feedback} 
		\label{alg:FeedbackDec_algorithm}
		\textbf{Input}:$\mathbf{x} =\textrm{sign} (\textbf{y}) $, $H$ (parity-check matrix), and maximum iteration number.
		
		\textbf{Output}: estimated codeword $\hat{\mathbf{c}}$.
		
		\begin{algorithmic}[1]
			\State $\mathbf{x}_{\textrm{in}} = {(1-\mathbf{x})}/{2}$
			\State $\mathbf{s}_{\textrm{in}} = \mod{(\mathbf{x}_{\textrm{in}} H^T, 2)}$
			\While{$\mathbf{s}_{\textrm{in}} \neq 0$ and maximum iterations is not reached}
			\State $\mathbf{x}_{\textrm{out}} = \Phi(\mathbf{x}_{\textrm{in}})$     \Comment{output of the chosen decoder}
			\State $\textbf{s}_{\textrm{out}} = \mod{(\mathbf{x}_{\textrm{out}} H^T, 2)}$
			\If{$\textbf{s}_{\textrm{out}} \neq 0$}
			\State $\mathbf{a} = \pi^{\star}(\mathbf{s}_{\textrm{in}})$    \Comment{Output of RL Block}
			\State $\mathbf{x}_{\textrm{in}} \gets \mathbf{x}_{\textrm{in}} \oplus \mathbf{a} $
			\State $\mathbf{s}_{\textrm{in}} = \mod{(\mathbf{x}_{\textrm{in}} H^T, 2)}$
			\State $\hat{\mathbf{c}} = \textbf{x}_{\textrm{in}}$
			\Else
			\State $\hat{\mathbf{c}} = \textbf{x}_{\textrm{out}}$
			\State Break
			\EndIf
			\EndWhile
		\end{algorithmic}
	\end{algorithm}
	
	\begin{theorem}
		The error correction capability of the proposed feedback decoder, utilizing the reward function defined in Equation (\ref{eq:reward_feedbackDec}), is lower bounded by
		\begin{equation}
			\lfloor \dfrac{\min_{y \in \mathcal{R}^{\Phi}_{f}}{w_{H}(y) } + \min_{y \in \mathcal{R}^{\Phi}_{m}}{w_{H}(y)} - 1}{2} \rfloor
		\end{equation}
		where $w_H(\cdot)$ denotes the Hamming weight of an error.
		\label{th::FD}
	\end{theorem}
	
	\begin{proof}
		The proof is provided in the Appendix \ref{Appendix:proof2}.
	\end{proof}

	\begin{remark}
		\label{remark:1}
		To correct the entire failure region, it is essential to have precise knowledge of the miscorrection region. Accordingly, the reward function is modified as follows
		\begin{equation}
			R(s,a,s^{\prime}) = 
			\begin{cases}
				-\frac{1}{L} +1, \quad & \textrm{if } s^{\prime} = S^{\Phi}_{c} \\
				-\frac{1}{L},   & \textrm{if } s^{\prime} \in S^{\Phi}_f \\
				-\frac{1}{L} -1, & \textrm{if } s^{\prime} \in S^{\Phi}_m
			\end{cases}
			\label{eq::FDMR}
		\end{equation}
		Then, the error-correction capability of the integrated decoder, after the convergence of training, is lower bounded by the minimum of the weight of the miscorrected error and the error-correction capability of the code, i.e.,
		\begin{equation}
			\min \left( t, \min_{y \in \mathcal{R}^{\Phi}_{m}}{w_H(y)} \right).
		\end{equation}
	\end{remark}
	To further reduce the number of states, we employ the concept of truncated MDP to decrease the state space size in the following Section similar to Section \ref{sec:TruncMDP}.
	
	\subsection{Feedback Decoder with Truncated MDP}
	\label{subsec::FBTItDec}
	By combining the concept of the feedback decoder presented in Section \ref{sec:Feedback_dec} with the truncated Markov decision process (MDP) discussed in Section \ref{sec:TruncMDP}, the number of states can be further reduced, thereby facilitating the training process and enabling the identification of the optimal policy. We define $\mathcal{BR}^{\Phi}_{c} (w) $ as the bounded successful decoding region of a given decoder $\Phi$ with radius $w$, represented as
	\begin{equation}
		\mathcal{BR}^{\Phi}_{c} (w) = \mathcal{B}(w) \cap \mathcal{R}^{\Phi}_{c}.
	\end{equation}
	Similarly, we define the bounded failure and miscorrection decoding regions of the decoder $\Phi$ as
	\begin{align}
		\mathcal{BR}^{\Phi}_{f} (w) &= \mathcal{B}(w) \cap \mathcal{R}^{\Phi}_{f} \\
		\mathcal{BR}^{\Phi}_{m} (w) &= \mathcal{B}(w) \cap \mathcal{R}^{\Phi}_{m}.
	\end{align}
	We denote ${BS}^{\Phi}_i (w)$ as the set of syndromes corresponding to the $i$ decoding region $\mathcal{BR}^{\Phi}_{i} (w)$ of the decoder $\Phi$, where $i \in \{c,f,m\}$ represents the correctable, failure, and miscorrectable region, respectively. ${BS}^{\Phi}_i (w)$ is defined as
	\begin{equation}
		{BS}^{\Phi}_i (w) = \{\mathbf{y} H^T \mid \mathbf{y} \in \mathcal{BR}^{\Phi}_{i} (w) \}.
	\end{equation}
	The action space is defined as in pervious sections, i.e., $ A = \{1,2,\cdots,n\}$. The reward function, similar to Equation \ref{eq:reward_feedbackDec}, is defined as
	\begin{equation}
		R(s,a,s^{\prime}) = 
		\begin{cases}
			-\frac{1}{L} +1, \quad & \textrm{if } s^{\prime} \in \overline{BS}^{\Phi}_f (w) \cap S(w)  \\
			-\frac{1}{L},   & \textrm{is } s^{\prime} \in {BS}^{\Phi}_f (w) \\
			-\frac{1}{L} -1, & \textrm{if } s^{\prime} \in \overline{S(w)}
		\end{cases}
		\label{eq::BF}
	\end{equation}
	The error correction capability of bounded feedback decoder, after the convergence of training with the reward function defined in Equation \ref{eq::BF} and based on Theorem \ref{th::FD} is given by
	\begin{equation}
		\min \bigg(  \bigg\lfloor \dfrac{\min_{y \in \mathcal{BR}^{\Phi}_{f} (w)}{w(y) } + \min_{y \in \mathcal{BR}^{\Phi}_{m} (w)}{w(y)} - 1}{2} \bigg\rfloor, w \bigg)
	\end{equation}
	where the minimum of the empty set is considered to be infinity, i.e., $\min {\emptyset} = \infty$.
	We can also define a reward function similar to Equation \ref{eq::FDMR} as
	\begin{equation}
		R(s,a,s^{\prime}) = 
		\begin{cases}
			-\frac{1}{L} +1, \quad & \textrm{if } s^{\prime} = {BS}^{\Phi}_{c}(w) \\
			-\frac{1}{L},   & \textrm{if } s^{\prime} \in {BS}^{\Phi}_f (w) \\
			-\frac{1}{L} -1, & \textrm{if } s^{\prime} \in {BS}^{\Phi}_m (w) \cup \overline{S(w)}.
		\end{cases}
		\label{eq::BFM}
	\end{equation}
	The error correction capability of bounded feedback decoder, after the convergence of training with the reward function defined in Equation \ref{eq::BFM} and based on Remark \ref{remark:1}, is given by
	\begin{equation}
		\min \big(\min_{y \in \mathcal{BR}^{\Phi}_{m} (w)} w(y) -1 , w \big).
	\end{equation}
	
	To further simplify the state space, in the next section, we focus on the automorphism group of group-structured quasi-cyclic LDPC codes.

	
	\section{Group-Structured Quasi-Cyclic LDPC Codes}
	\label{sec:automorphism_QC-LDPC}
	Quasi-Cyclic Low Density Parity Check (QC-LDPC) codes play a crucial role in practical settings due to their structured design. Unlike randomly constructed LDPC codes, a quasi-cyclic LDPC is specified by a parity-check matrix $H$ composed of circulant submatrices. Specifically, $H$ is structured as a $j \times k$ array of $p \times p$ circulant permutation matrices, denoted $H_{s,t}$ for $0 \leq s < j$ and $0 \leq t < k$, where $p$ is prime. The integers $a, b \in \mathbb{F}_p$ and $a \neq b$ with orders $k $ and $j$, respectively, which in turn requires $k$ and $j$ to be divisors of $p-1$. Each circulant matrix $H_{s,t}$ can be viewed as an identity matrix cyclically shifted by the amount $P_{s,t} = b^{s} a^{t}$. The matrix $H$ is a $(jp) \times (kp)$ matrix whose each row contain exactly $k$ ones and each column contain exactly $j$ ones, making it a regular $(j,k)$ LDPC code \cite{tanner2001class}.

	
	\subsection{Automorphism Group}
	Since the optimal policy may be learned for a given syndrome but not for its isomorphic counterpart, the diversity introduced by automorphic mappings can be leveraged to improve decoding performance. An automorphism of a bipartite graph is a permutation of the vertices (i.e., variable nodes and check nodes) that preserves adjacency. In other words, it is a bijection that maps variable nodes to variable nodes and check nodes to check nodes while ensuring that edges remains invariant. The cyclic shift automorphism arises by cyclically shifting each coordinate modulo $p$. Specifically, if one adds a fixed integer $\delta \; (\textrm{mod}) \; p $ to a coordinate $x_i$, the value cycle through $\{ 0, \dots, p-1 \}$. Since each row block, or circulant block, of $H$ just shifts its columns modulo $p$, these additive shifts do not change adjacency structure. Consequently, there exists a group of automorphisms isomorphic to the additive group $\mathbb{Z}_p$. We consider two multiplicative subgroups. Define an automorphism $\pi$ on the set of variable nodes by $\pi: \;  (x_1, x_2, \dots, x_p) \mapsto (a x_1, a x_2, \dots, a x_p) \;\; (\textrm{mod} \; p) ,$ where each $x_i \in \{0, 1, \dots, p-1 \}$ and all indexes computed mod $p$. Automorphism $\pi$ maps the bits of the $t$-th block to the bits of the $(t +1 \mod k)$-th block. Similarly, an automorphism $\rho$ can be defined that maps the block of parity checks to the next block of parity checks by $\rho: \;  (x_1, x_2, \dots, x_p) \mapsto (b x_1, b x_2, \dots, b x_p) \;\; (\textrm{mod} \; p) ,$ where each $x_i \in \{0, 1, \dots, p-1 \}$ and all indexes computed mod $p$. These automorphisms, both additive and multiplicative, exploit the inherent symmetry of the quasi-cyclic structure and ensure that the overall adjacency in the Tanner graph of the code is preserved. Let $\mathbf{e}$ denote the hard-decision received vector, and let $\mathcal{S}$ be the syndrome operator that computes the syndrome $\mathbf{s}$ associated with $\mathbf{e}$. Based on this, we define the following automorphism-based strategy
	\begin{equation}
		\scalebox{0.7}{
			$\begin{array}{ccc}
				e_1 & \xrightarrow{\;\mathcal{S}\;} & s_1 \\[6pt]
				\Gamma_{v}^{-1}\bigg\uparrow\bigg\downarrow{\Gamma_{v}} & & \Gamma_{s}^{-1}\bigg\uparrow\bigg\downarrow{\Gamma_{s}} \\[6pt]
				e_2 & \xrightarrow{\;\mathcal{S}\;} & s_2
			\end{array}$
		}
	\end{equation}
	where $v_2 = \Gamma_{v}(v_1)$, $s_2 = \Gamma_{s}(s_1)$, $S(v_1) = s_1$, and $S(v_2) = s_2$. This expression illustrates the property that the syndrome mapping and the automorphism $\Gamma$ commute. This indicates that the syndrome mapping and the automorphism commute.
	\begin{equation}
		\mathcal{S}(\Gamma_{v}(e)) = \Gamma_{s}(\mathcal{S}(v))
	\end{equation}
	This property is significant because it implies that the automorphism $\Gamma$ preserves the syndrome structure of the code. It shows that the symmetry given by the quasi-cyclic structure extends naturally to both the codeword space and the syndrome space. 

	
	
	
	\subsection{Number of Unique Structures}
This section focuses on quantifying the number of unique states in the QC-LDPC code to analyze the reduction in state space and its impact on decoder performance. We first formulate the counting problem to determine the number of unique states under the symmetry groups introduced in the previous section. Subsequently, we define the notion of the smallest lexicographic ordering, which allows us to represent each equivalence class of states by a single canonical representative. This enables us to consider only one unique state per equivalence class in the MDP. The necessary group theory background is provided in Appendix \ref{Appendix:GroupTheory}. Let $p$ be a prime number, and let $b\in(\mathbb{Z}/p\mathbb{Z})^\times$ be an element of multiplicative order exactly $j$, i.e., $b^j\equiv1\pmod p$. We consider $j$ disjoint necklaces, each consisting of $p$ beads. We label each necklace by an index $s \in \mathbb{Z}/j \mathbb{Z}$. We then define the bead set as
\begin{equation}
	\Omega \;=\;\{(s,i)\;:\;s\in\mathbb{Z}/j \mathbb{Z},\;i\in\mathbb{Z}/p \mathbb{Z}\}
\end{equation}
where $(s,i)$ denotes the $i$-th bead of the $s$-th necklace. We define two symmetry group actions on the bead set $\Omega$. The first is a rotation within each necklace, given by
\begin{equation}
	\sigma(s,i) = (s,\,i+1\pmod p),
\end{equation}
which corresponds to a cyclic shift of the beads within the $s$-th necklace. The second is a multiplicative action defined by
\begin{equation}
	\rho(s,i) = (\,s+1 \mod j,\;b\,i \mod p).
\end{equation}
This transformation simultaneously advances to the next necklace and permutes the bead positions through multiplication by $b$. Thus, the symmetry group is given by the semidirect product
\begin{equation}
	G \;=\;\langle\sigma,\rho \mid \sigma^p=1,\;\rho^j=1,\;\rho\,\sigma\,\rho^{-1}=\sigma^b\rangle
	\;\cong\;C_p\!\rtimes C_j.
\end{equation}
We define the set of all binary colorings of the bead set $\Omega$ as
\begin{equation}
	X \;= \{0, 1\}^\Omega =\;\{\,f:\Omega\to\{0,1\}\}
\end{equation}
where each function $f \in X$ assigns a binary color $0/1$ to each bead in $\Omega$. We define an action of the group $G$ on the set $X$ by 
\begin{equation}
	(g\cdot f)(\omega) \;=\; f\bigl(g^{-1}(\omega)\bigr),
	\quad
	\forall g\in G,\; \forall \omega\in\Omega,
\end{equation}
where $f\in X$. This is the left group action on functions, induced by the left action of $G$ on $\Omega$. Two colorings $f$ and $f^\prime$ are equivalent if there exists a group element $g \in G$ such that $f^\prime = g \cdot f$. Our goal is to determine the set of orbits under the group action, which corresponds to the set of distinct colorings up to symmetry
\begin{equation}
	X/G \;=\;\{\,[f]\mid f\in X\},
\end{equation}
where $[f]$ denotes the equivalence class of $f \in X$ under the action of $G$. To compute the cardinality of $X/G$, we apply Burnside's Lemma
\begin{equation}
	\begin{split}
		\bigl\lvert X/G\bigr\rvert
		& = \frac{1}{\lvert G\rvert}
		\sum_{g\in G}
		\bigl\lvert\{\,f\in X : g\cdot f = f\}\bigr\rvert \\
		& = \frac{1}{jp}
		\sum_{g\in G}
		2^{\#\text{cycles of }g\text{ on }\Omega}.
	\end{split}
\end{equation}
where the number of fixed points of $g$ acting on $X$ is equal to $2^c$, with $c$ denoting the number of cycles of $g$ in its action on the domain $\Omega$. 

\begin{theorem}
	The number of unique syndromes, $N_s$, in the described QC-LDPC code is bounded by
	\begin{align}
		&\frac{1}{2^{m-\textrm{rank}(H)}} N_{\textrm{full}} \leq N_s \leq N_{\textrm{full}} \nonumber \\
		&N_{\textrm{full}} = \frac{1}{jp}\Bigl[\, 2^{jp} + (p-1)\,2^j + p\sum_{\substack{d\mid j\\d<j}}\varphi\!\bigl(\tfrac{j}{d}\bigr)\,2^{p\,d} \Bigr],
	\end{align}
	where $j$ and $p$ define the QC structure, and $\varphi (\cdot)$ denotes the Euler's totient function.
\end{theorem}

\begin{proof}
	The proof is provided in the Appendix \ref{Appendix:proof3}.
\end{proof}

\subsection{Canonical Lexicographic Representation}
To eliminate redundancy arising from symmetry, we use a single representative for each orbit of colorings. Selecting the lexicographically smallest element within each orbit as its canonical representative. Let $\Omega$ denotes the set of bead positions, with cardinality $N$, and let $f$ be the binary coloring function. We fix once for all beads a bijection $\pi\colon \{1,2,\dots,N\}\;\longrightarrow\;\Omega$. This identification allows us to associate each coloring $f$ with a bit-vector
\begin{equation}
	\mathbf{v}_f \;=\;\bigl(f(\pi(1)),\,f(\pi(2)),\dots,f(\pi(N))\bigr)\;\in\;\{0,1\}^N.
\end{equation}
We compare two vectors $\mathbf{v}, \mathbf{v}^{\prime} \in\{0,1\}^N$ using the {\em lexicographic order}, defined as follows in
\begin{equation}
	\mathbf{v} <_{\rm lex} \mathbf{v}^{\prime} 
	\; \Longleftrightarrow \;
	\exists\,k \colon \; \mathbf{v}_i = \mathbf{v}^{\prime}_i\;(i<k), \;
	\mathbf{v}_k < \mathbf{v}^{\prime}_k.
\end{equation}
In other words, $\mathbf{v}$ precedes $\mathbf{v}^{\prime}$ in lexicographic order if the first position at which they differ has $\mathbf{v}_k < \mathbf{v}^{\prime}_k$. We select the unique lexicographically smallest element within each orbit.
\noindent\textbf{Definition.}
The {\em canonical representative} of the orbit of a coloring $f$ is
\begin{equation}
	\mathrm{can}(f) = \arg\min_{g\in G}\;v_{\,g\cdot f} \;\in\;\{0,1\}^N,
\end{equation}
that is, the coloring in the orbit $G\cdot f$ whose associated bit vector is minimal with respect to lexicographic order. Based on Booth’s algorithm \cite{BOOTH1980240}, the lexicographically minimal rotation of a length-$N$ vector can be found in $O(N)$. Since there are $j$ coset representatives arising from the multiplicative group action, the overall computational cost to determine the minimal representative of an orbit is $O(jN) = O(j^2 p)$.

\section{Numerical Results}
\label{sec:num_result}
\begin{figure*}
	\centering
	\begin{minipage}[t]{0.5\textwidth}
		\centering
		\includegraphics[width=\textwidth]{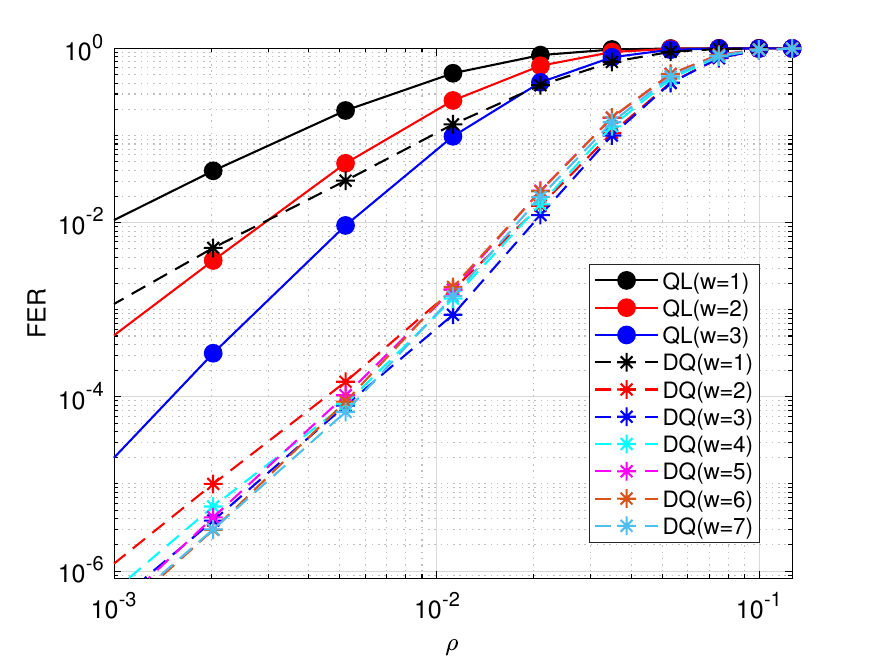}
		\caption*{(a)}
		\label{fig:FER_RLTanner}
	\end{minipage}%
	\hfill
	\begin{minipage}[t]{0.5\textwidth}
		\centering
		\includegraphics[width=\textwidth]{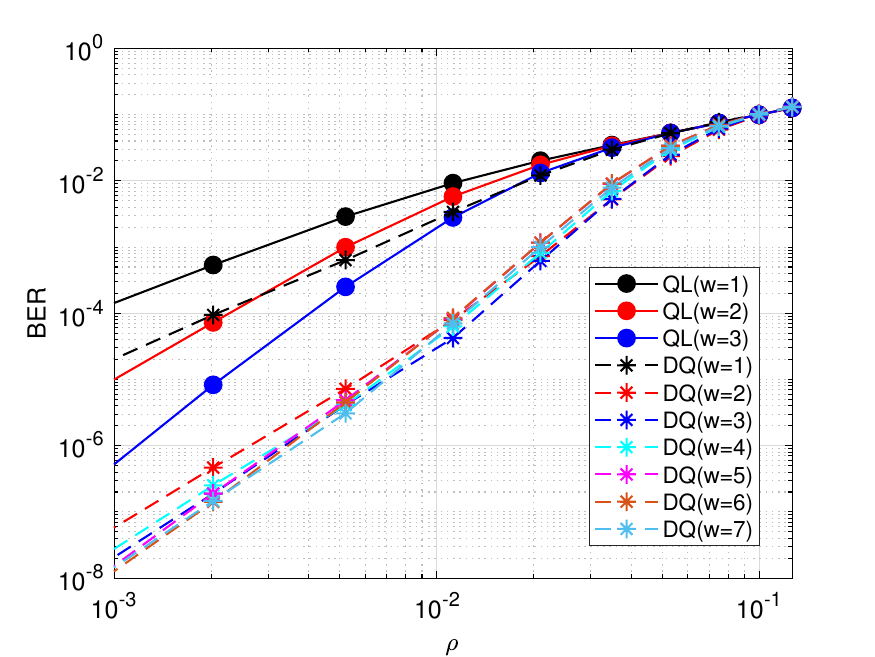}
		\caption*{(b)}
		\label{fig:three sin x}
	\end{minipage}
	\caption{Simulation results for tanner code of Bit-Flipping decoder and learned Bit-Flipping based on Q-table for truncated MDPs with weight $w=\{1,2,3 \}$ and Deep-Q learning with weights $w=\{1,2,\cdots,7 \}$. (a) Frame Error Rate, (b) Bit Error Rate. 
	}
	\label{fig:tanner_Q_DeepQ}
\end{figure*}

\begin{figure*}
	\centering
	\begin{minipage}[t]{0.5\textwidth}
		\centering
		\includegraphics[width=\textwidth]{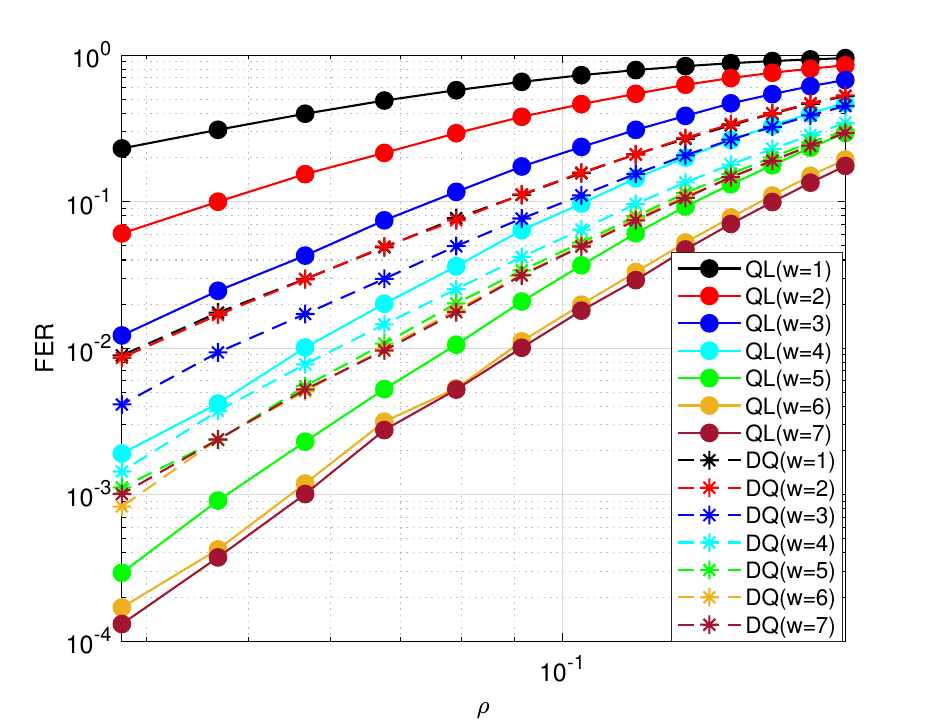}
		\caption*{(a)}
		\label{fig:y equals x}
	\end{minipage}%
	\hfill
	\begin{minipage}[t]{0.5\textwidth}
		\centering
		\includegraphics[width=\textwidth]{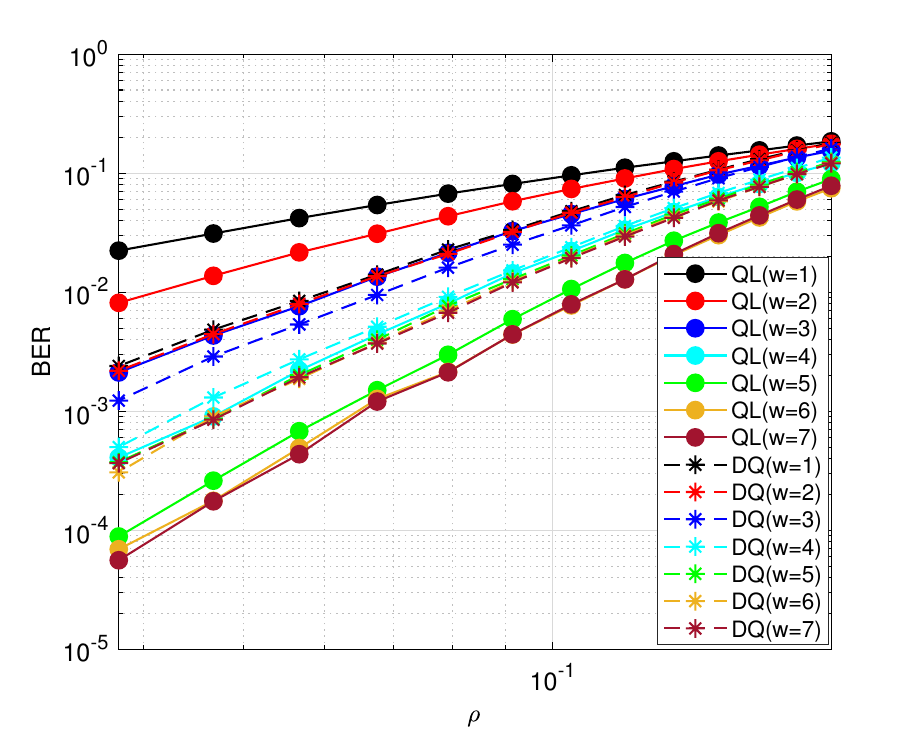}
		\caption*{(b)}
		\label{fig:three sin x}
	\end{minipage}
	\caption{Simulation results for a binary random linear code with parameters ($24,6,10$) of Bit-Flipping decoder and learned Bit-Flipping based on Q-table for truncated MDPs with weight $w=\{1,2,3,4,5,6,7 \}$. (a) Frame Error Rate, (b) Bit Error Rate.}
	\label{fig:n24}
\end{figure*}

\begin{figure*}
	\centering
	\begin{minipage}[t]{0.50\textwidth}
		\centering
		\includegraphics[width=\textwidth]{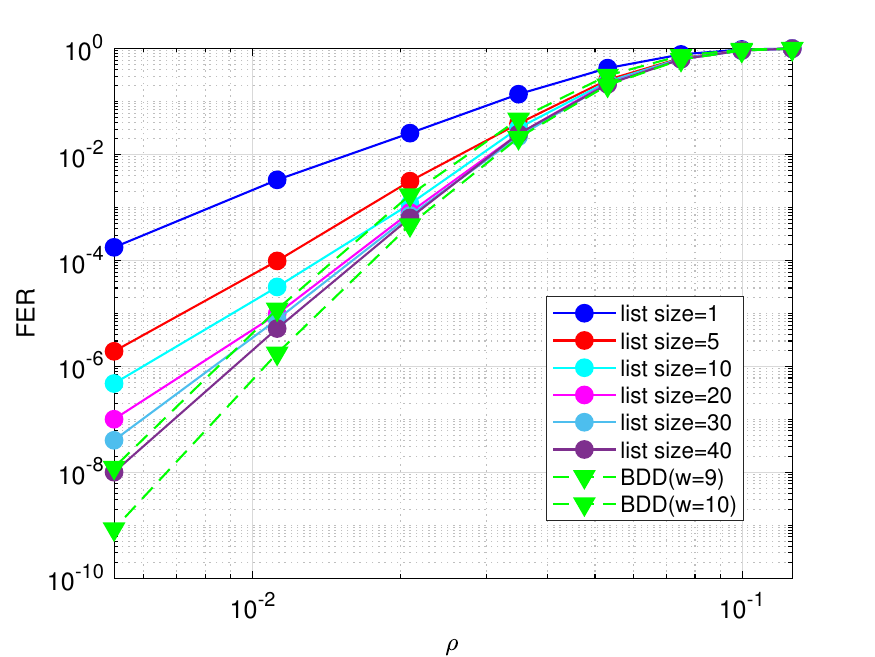}
		\caption*{(a)}
		\label{fig:FER_list_decoding_RLTanner}
	\end{minipage}%
	\hfill
	\begin{minipage}[t]{0.50\textwidth}
		\centering
		\includegraphics[width=\textwidth]{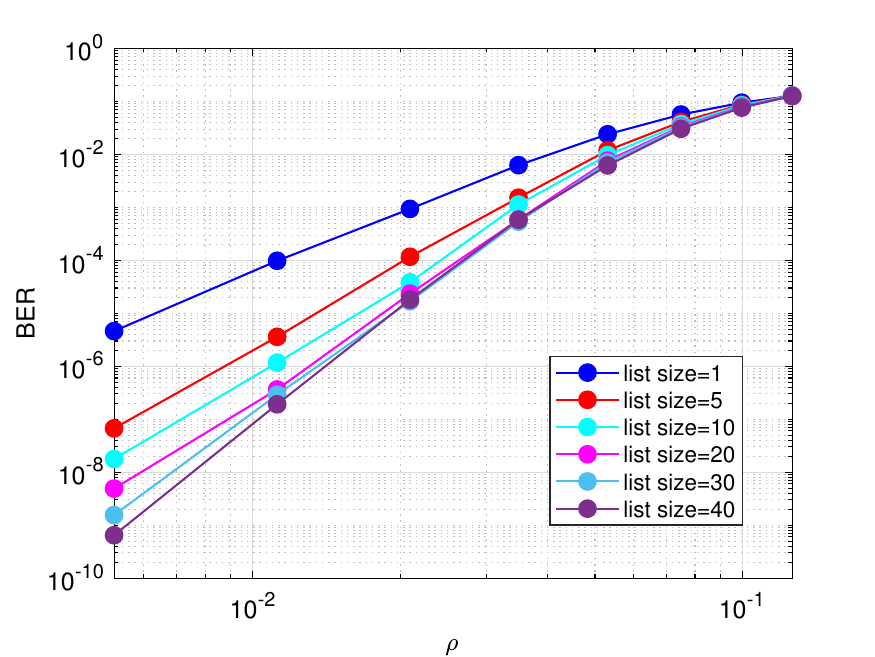}
		\caption*{(b)}
		\label{fig:BER_list_decoding_RLTanner}
	\end{minipage}
	\caption{Simulation results for Tanner code with Deep Q-Network decoder and different beam sizes $\{1, 5, 10, 20, 30, 40\}$.\\ (a) Frame Error Rate, (b) Bit Error Rate.}
	\label{fig:list_decoding_RLTanner}
\end{figure*}

\begin{figure*}
	\centering
	\begin{minipage}[t]{0.50\textwidth}
		\centering
		\includegraphics[width=\textwidth]{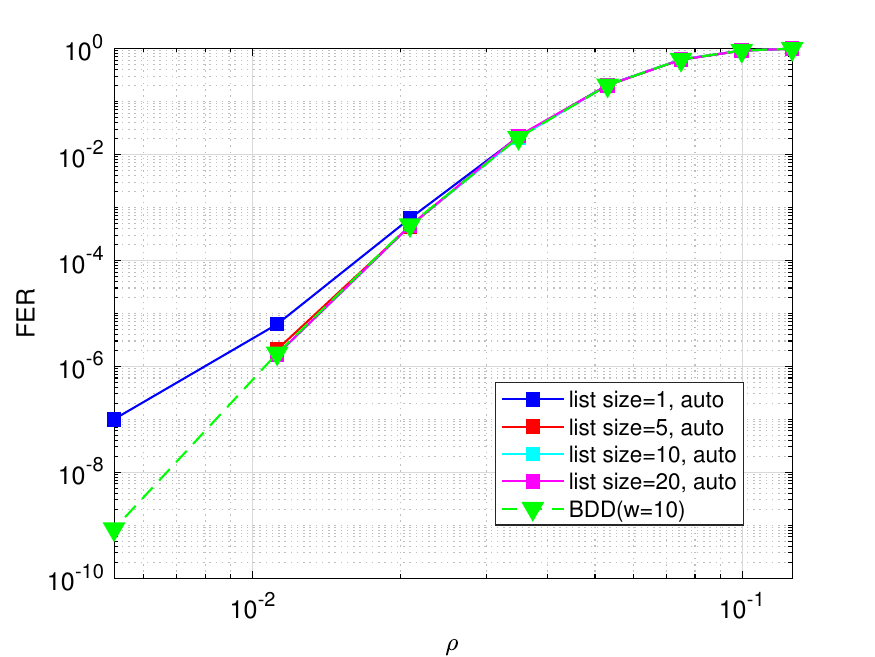}
		\caption*{(a)}
		\label{fig:FER_list_decoding_automorphism_RLTanner}
	\end{minipage}%
	\hfill
	\begin{minipage}[t]{0.50\textwidth}
		\centering
		\includegraphics[width=\textwidth]{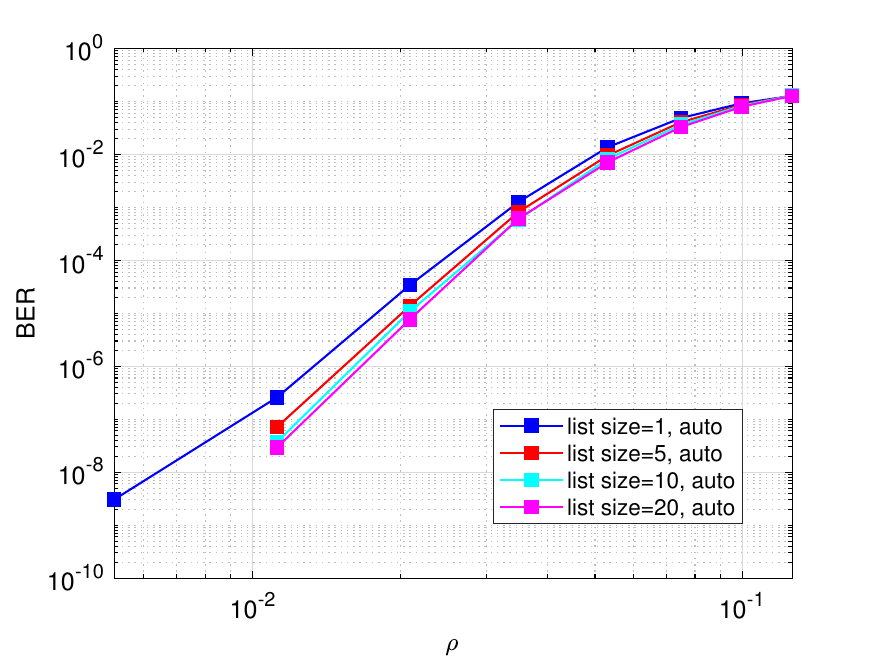}
		\caption*{(b)}
		\label{fig:three sin x}
	\end{minipage}
	\caption{Simulation results for Tanner code with Deep Q-Network decoder and different beam sizes $\{1, 5, 10, 20\}$ with \\ cyclic permutation strategy. (a) Frame Error Rate, (b) Bit Error Rate.}
	\label{fig:list_decoding_automorphism_RLTanner}
\end{figure*}

\begin{figure*}
	\centering
	\begin{minipage}[t]{0.50\textwidth}
		\centering
		\includegraphics[width=\textwidth]{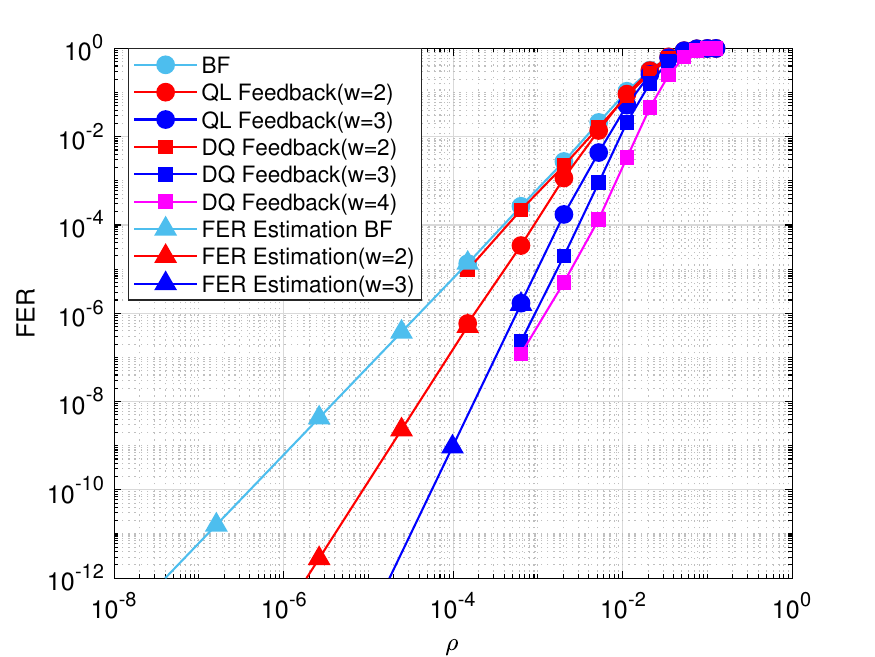}
		\caption*{(a)}
		\label{fig:FER_RLTanner_RLFeedback}
	\end{minipage}%
	\hfill
	\begin{minipage}[t]{0.50\textwidth}
		\centering
		\includegraphics[width=\textwidth]{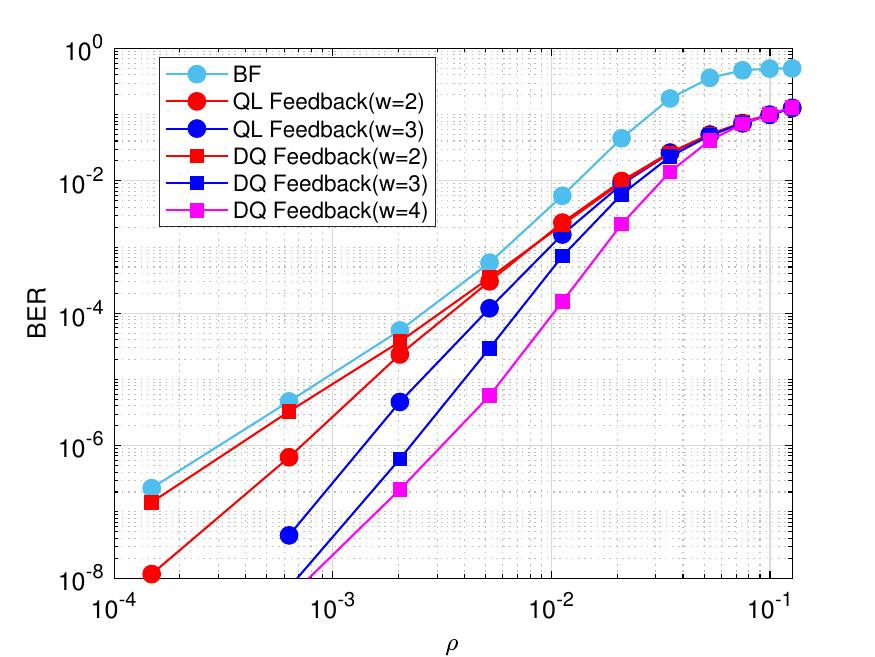}
		\caption*{(b)}
		\label{fig:three sin x}
	\end{minipage}
	\caption{Simulation results for the Tanner code using the Bit-Flipping decoder and and its Q-learning and Deep-Q network feedback methods for truncated MDPs with radii $w=\{2,3,4 \}$. (a) Frame Error Rate, (b) Bit Error Rate.}
	\label{fig:tanner_Q_DeepQ_RLFeedback_BF}
\end{figure*}

In this section, we present numerical results for the Tanner code with parameters $(155, 64, 20)$ for the truncated learned bit-flipping decoder, action-list decoder, the action-list decoder with code automorphism, and the feedback decoder, as described in Sections \ref{sec:TruncMDP}, \ref{sec:beam-decoder}, \ref{sec:Feedback_dec}, and \ref{sec:automorphism_QC-LDPC}, respectively. In Figure \ref{fig:tanner_Q_DeepQ}, we present the performance of the Tanner code, which is a low-rate, moderate-length code. The truncated MDP described in Section \ref{sec:TruncMDP} is employed to determine the optimal flipping rule using the Q-learning algorithm. We consider three different hamming ball radii, $w=\{1,2,3\}$, which result in the number of states in the truncated MDP being $156$, $12091$, and $620776$, respectively. For $w=4$ and beyond, using a Q-table becomes infeasible due to the exponentially large state space. The performance of the bounded-distance decoding (BDD) method, described in Equation \ref{eq::BDD_error}, aligns with the performance of the Q-learning method. This serves as a baseline to assess whether the Q-learning algorithm has successfully converged. Additionally, we employ the Deep Q-learning algorithm under a truncated MDP with different Hamming ball radii, $w=\{1,\cdots,7\}$. The results indicate that for weights $w=\{1,2,3\}$, even though the agent was only exposed to error failures related to weight-$1$, weight-$2$, or weight-$3$ error patterns during training, it was able to generalize and correct some states associated with higher-weight error patterns that it had never experienced during training. After a certain point, exposing the agent to additional states for learning the optimal policy yields diminishing benefits. This is primarily due to the limited capacity of the neural network, which may not be sufficiently expressive to fully capture the complexity of the expanded state space and the corresponding reward signal values.
In Figure \ref{fig:n24}, the results illustrate a similar phenomenon for a short-length and low-rate code, providing further clarification. We select a random linear block code of length $24$ with a minimum distance of $10$ that does not have a specific structure. The performance of the Q-learning and Deep Q-learning algorithms is simulated for different Hamming ball radii, $w=\{1, \cdots, 7\}$. The results indicate that, for a random code, Deep Q-learning is capable of generalizing the optimal policy for states that it has never encountered during training for $w=\{1, \cdots, 4 \}$. However, for radii beyond $4$, the neural network's capacity to approximate Q-values becomes insufficient to fully capture the complexity of the significantly larger state space and the corresponding reward signal values. It is worth mentioning that, since the code is not a perfect code, we extend the Hamming ball radius to $7$.
In Figure \ref{fig:list_decoding_RLTanner}, we evaluate the proposed Q-value guided action-list decoder, introduced in Section \ref{sec:beam-decoder}, using different list sizes $\{1, 5, 10, 20, 30, 40 \}$ to assess their impact on decoding performance. The results demonstrate a substantial improvement in performance as the number of candidate bit-flips increases, which helps the decoder avoid getting trapped in oscillatory behavior. For comparison, we also include the performance of bounded distance decoding (BDD) with radii $9$ and $10$ as baselines. In Figure \ref{fig:list_decoding_automorphism_RLTanner}, we present the performance of the action-list decoder with varying list sizes, incorporating the automorphism group of the Tanner code as described in Section \ref{sec:automorphism_QC-LDPC}. We only employ $(p-1)$ cyclic permutations from the automorphism group of the tanner code. As shown in Figure \ref{fig:list_decoding_automorphism_RLTanner}, the performance of the action-list decoder with a list size of $5$, combined with the permutation group, closely approaches that of bounded distance decoding (BDD) with radius $10$. We additionally present the performance of the action-list decoder with list sizes $\{10, 20\}$, incorporating automorphisms. While the frame error rate (FER) performance remains similar through Monte Carlo simulation, the bit error rate (BER) exhibits a slight improvement. 
For the proposed feedback decoder described in Section \ref{sec:Feedback_dec}, we select the arbitrary decoder $\Phi$ to be the bit-flipping decoder, which, in general, can be any well-established code and decoder pair from the literature. The goal is to enhance its performance through the integration of the RL block. The bounded error weight enumerator of the Tanner code under the BF decoder, around all-zero codeword, is $ E(w=4) = 10 092 825 x^4 + 154225 x^3 + 620 x^2$, where $w$ denotes the maximum weight of the error patterns considered during decoding. The multiplicity of $x^i$ represents the number of uncorrectable weight-$i$ errors under BF decoder. In addition, no miscorrections are observed for this code when employing the BF decoder for error patterns with weights of at most $4$. In Figure \ref{fig:tanner_Q_DeepQ_RLFeedback_BF}, we illustrate the performance of the bit-flipping algorithm alongside truncated MDPs with radii $w = \{2, 3 \}$, trained using the Q-learning, and radii $w=\{2,3, 4\}$, trained using the Deep Q-learning algorithms. We also present the frame error rate (FER) estimates in the error floor region for the feedback model learned using the Q-learning algorithm based on their uncorrectable error patterns. The results demonstrate that, instead of training the RL block to learn the entire code space, we can use the existing bit-flipping algorithm and incorporate the RL block as a feedback architecture to enhance its performance. The Q-learning algorithm with $w=\{2, 3\}$ achieves improved performance with guaranteed error correction. However, for $w=4$, the state space becomes prohibitively large, making it impractical to use a Q-table. In this case, we train only the Deep Q-Network. Using the Deep-Q network, for $w=2$, it does not effectively learn the decision boundary due to the small radius of the Hamming ball. However, for $w=3$, as shown by comparison with the Q-table results, the network is able to generalize the optimal policy and successfully correct error patterns were not encountered during training. For $w=4$, further performance improvement is observed in comparison to $w=3$. 

\subsection{Training Hyper-Parameters}
For training the Deep Q-learning algorithm, we employ a neural network with one hidden layer consisting of $512$ neurons. The maximum number of decoding iterations is set to $D_{\max} = 10$. The training process involves $10^{6}$ episodes, with a batch size of $128$. For the exploration-exploitation strategy, an epsilon-greedy approach with linear decay is employed, where $\epsilon$ decreases from $0.9$ to $0.05$. The learning rate is set to $10^{-4}$. In Figures \ref{fig:list_decoding_RLTanner}, \ref{fig:n24}, and \ref{fig:list_decoding_automorphism_RLTanner}, the discount factor is set to $\gamma = 0.9$, whereas in Figures \ref{fig:tanner_Q_DeepQ} and \ref{fig:tanner_Q_DeepQ_RLFeedback_BF}, the discount factor is $\gamma = 0.7$. In Figures \ref{fig:n24}, we employ a neural network with a single hidden layer consisting of $128$ neurons.

\section{Conclusions}
\label{sec:Conclusions}
In this paper, we propose a novel action-list decoding strategy aimed at enhancing decoding performance by considering a list of high-probability error patterns, rather than relying on greedy action selection during the inference stage. We, also, propose a novel feedback method aimed at reducing the number of states in the MDP. Additionally, we focus on enhancing the performance of a chosen decoder by targeting only those error patterns that the decoder fails to correct. In essence, the feedback method seeks to efficiently expand the correctable region of the decoder. In general, our goal is to leverage the performance of existing well-designed decoders and refine their state space and using RL algorithms to further enhance their decoding performance. Additionally, by leveraging the automorphism group, we further improve the effectiveness of the proposed method within the region explored through Monte Carlo simulations.

\section*{Acknowledgments}
The authors acknowledge the support of the National Science Foundation under grants CIF-2420424, CIF-2106189, CCF-2100013, CCSS-2052751, and CCSS-2027844. Bane Vasi\'c has disclosed an outside interest in his company Codelucida to The University of Arizona. Conflicts of interest resulting from this interest are being managed by The University of Arizona in accordance with its policies.


\bibliographystyle{IEEEtran}
\bibliography{bib/references}

\appendices
\section{Proof of Theorem 1}
\setcounter{theorem}{0}
\label{Appendix:proof1}

\begin{theorem}
	The number of distinct optimal policies, or proposed decoders, for the Binary Symmetric channel is at least
	\begin{equation}
		\prod_{i=1}^{t} i^{\binom{n}{i}} \nonumber
		\label{eq:nopt_pol}
	\end{equation}
	where $n$ denotes the codelength and $t$ represents the error-correcting capability of the code.
\end{theorem}

\begin{proof}
	The total number of distinct syndromes is $2^{n-k}$. Let assume that the code has the error correction capability of $t$. In this case, all error patterns with a weight less than or equal to $t$ are correctable. There exists a one-to-one correspondence between syndromes and their associated coset leader, satisfying $\sum_{i=1}^{t} \binom{n}{i} \leq 2^{n-k}$, which implies the possibility of correcting certain error patterns with a weight greater than $t$. There are $i$ positions in a weight-$i$ error pattern that can be flipped to correctly transition to a weight-$(i-1)$ error pattern, where $ 1 \leq i \leq t$.
	For each weight-$i$ error pattern, there exist $\binom{n}{i}$ distinct syndromes, each with $i$ optimal actions. Thus, the number of possible optimal actions transitioning from a weight-$i$ error pattern to a weight-$(i-1)$ error pattern is is $i^{\binom{n}{i}}$, where $ 1 \leq i \leq t$.
	Since the correction of error patterns can be viewed as a sequential procedure, the multiplication axiom can be applied iteratively. We can state that the total number of distinct optimal policies, considering the correction of error patterns up to weight $t$, is determined by the product of the possible transitions from a weight-$i$ error pattern to a weight-$(i-1)$ error pattern for $1 \leq i \leq t$. Therefore, the number of distinct optimal policies up to correction of weight-$t$ error patterns is given by $ \prod_{i=1}^{t} i^{\binom{n}{i}} $. 
	If the code is perfect, the $t$-radius decoding region includes all syndromes, and thus the product above counts all optimal policies. Otherwise, there are additional correctable syndromes which coset leader has weight greater than $t$. These additional syndromes increase the number of optimal policies, making the product a lower bound.
\end{proof}

\section{Proof of Theorem 2}
\label{Appendix:proof2}
\setcounter{theorem}{1}
\begin{theorem}
    The error correction capability of the proposed feedback decoder, utilizing the reward function defined in Equation (\ref{eq:reward_feedbackDec}), is lower bounded by
    \begin{equation}
        \lfloor \dfrac{\min_{y \in \mathcal{R}^{\Phi}_{f}}{w_{H}(y) } + \min_{y \in \mathcal{R}^{\Phi}_{m}}{w_{H}(y)} - 1}{2} \rfloor \nonumber
    \end{equation}
    where $w_H(\cdot)$ denotes the Hamming weight of an error.
    \label{th::FD}
\end{theorem}

\begin{proof}
    The Reinforcement Learning (RL) objective, based on the reward function defined in Equation (\ref{eq:reward_feedbackDec}), is to minimize the number of bit flips required to achieve the all-zero syndrome. The integration of RL with the existing decoder $\Phi$ aims to determine the optimal sequence of actions necessary to reach the all-zero syndrome. 
    After training on the failures of the existing decoder $\Phi$, three outcomes are possible: the syndrome associated with $\mathbf{y}$ may map to the correctable region, the miscorrection region, or the failure region. Upon the convergence of training, the syndrome associated with $\mathbf{y}$ will map to either the correctable set $\mathcal{R}^{c}$ or the miscorrection $\mathcal{R}^{m}$ set as follows
    \begin{equation}
        \forall y \in \mathcal{R}^{\Phi}_{f} : \min_{y' \in \mathcal{R}^{\Phi}_{c}} d(y, y')   \underset{\mathcal{R}^{c}}{\overset{\mathcal{R}^{m}}{\gtrless}} \min_{y' \in \mathcal{R}^{\Phi}_{m}} d(y, y')
    \end{equation}
    This implies that $\mathbf{y}$ will map to the region closest to it. We aim to identify the minimum weight of a failure-inducing error that results in a miscorrection set, thereby establishing the guaranteed error-correction capability of the decoder, which can be expressed as the optimization problem in Equation \ref{eq:opt_guaranteed}
    \begin{equation}
        \begin{aligned}
            \min_{\mathbf{y} \in \mathcal{R}^{\Phi}_{f}} \quad & w_H(\mathbf{y}) \\
            \textrm{s.t.} \quad & \min_{\mathbf{y}^{\prime} \in \mathcal{R}^{\Phi}_{c}} d(\mathbf{y}, \mathbf{y}^{\prime}) - \min_{\mathbf{y}^{\prime\prime} \in \mathcal{R}^{\Phi}_{m}} d(\mathbf{y}, \mathbf{y}^{\prime\prime}) > 0.
        \end{aligned}
        \label{eq:opt_guaranteed}
    \end{equation}
    Thus, the boundary between the correctable and miscorrectable region after convergence is defined as
    \begin{equation}
        \quad \min_{\mathbf{y}^{\prime} \in \mathcal{R}^{\Phi}_{c}} d(\mathbf{y}, \mathbf{y}^{\prime}) = \min_{\mathbf{y}^{\prime\prime} \in \mathcal{R}^{\Phi}_{m}} d(\mathbf{y}, \mathbf{y}^{\prime\prime}) 
    \end{equation}
    For any failure word on the boundary, the probability of reaching either the correctable or miscorrectable region is equal.
    \\
    To identify the failure region that is guaranteed to be correctable, we first determine $r_c$, the minimum weight of error patterns for which all errors with that weight are correctable by the decoder $\Phi$. In other words,  $r_c$  represents the minimum radius of the Hamming ball centered at the zero codeword such that all errors within this ball are correctable by the decoder $\Phi$. Additionally, we determine $r_m$, the minimum weight of miscorrected errors.
    Thus, any decoder failure with a weight below the boundary defined by the radius
    $r_c + {(r_m - r_c)}/{2} = {(r_m + r_c)}/{2}$
    lies closer to the correctable region of \( \Phi \) and can therefore be corrected. Consequently, all error failures with a weight lower than ${(r_m + r_c)}/{2}$ are correctable. Since all errors within the Hamming ball of radius $r_c$ are correctable, $r_c$  is defined as
    \begin{equation}
        r_c = \min \left( \min_{y \in \mathcal{R}^{\Phi}_{f}}{w_H(y) }, \min_{y \in \mathcal{R}^{\Phi}_{m}}{w_H(y)} \right) - 1
    \end{equation}
    If we assume that $\min_{y \in \mathcal{R}^{\Phi}_{f}}{w_H(y) } \leq \min_{y \in \mathcal{R}^{\Phi}_{m}}{w_H(y)}$, therefore, any error with a weight smaller than
    \begin{equation}
        \lfloor \dfrac{\min_{y \in \mathcal{R}^{\Phi}_{f}}{w_H(y) } + \min_{y \in \mathcal{R}^{\Phi}_{m}}{w_H(y)} - 1}{2} \rfloor
    \end{equation}
    is correctable.

\end{proof}

\section{Group Theory Basics}
\label{Appendix:GroupTheory}

Let $G$ be a finite group, and $X$ a set. A \textbf{left action} of $G$ on $X$ is a map
\begin{equation}
    G \times X \longrightarrow X, \quad (g, x) \mapsto g \cdot x,
\end{equation}
satisfying the following properties:
\begin{enumerate}
    \item \textbf{Identity:} \( e \cdot x = x, \)   \( \forall x \in X \), where \( e \in G \) is the identity element.
    \item \textbf{Associativity:} \( (gh) \cdot x = g \cdot (h \cdot x), \)  \( \forall g, h \in G \) and \( \forall x \in X \).
\end{enumerate}

\begin{definition}{(\textbf{Orbit})}.
    The \emph{orbit} of an element $x \in X$ under the action of a group $G$ is defined as
    \begin{equation}
        \textrm{Orb} (x) = \{ g \cdot x : g \in G  \} \subseteq X.
    \end{equation}
    It consists of all elements of $X$ to which $x$ can be mapped through the action of elements of $G$.
\end{definition}

\begin{definition}{(\textbf{Stabilizer}).}
    The \emph{stabilizer} of an element $x \in X$ is the subgroup of $G$ consisting of all elements that fix $x$ under the group action, defined by
    \begin{equation}
        \textrm{Stab} (x) = \{ g \in G : g \cdot x = x  \} \leq G.
    \end{equation}
\end{definition}

The \emph{Orbit-Stabilizer} Theorem establishes a fundamental relationship between orbits and stabilizers under a group action.
\begin{theorem}{(\textbf{Orbit-Stabilizer Theorem})}
    For every $x \in X$, there exists a natural bijection
    \begin{equation}
        G / \textrm{Stab}(x) \leftrightarrow \textrm{Orb} (x).
    \end{equation}
    In particular, if $G$ is finite, then
    \begin{equation}
        \mid \textrm{Orb} (x) \mid = \dfrac{\mid G \mid}{\mid \textrm{Stab} (x) \mid}.
    \end{equation}
\end{theorem}

We now state \emph{Burnside's Lemma}, a fundamental result for counting the number of distinct orbits under a group action.
\begin{theorem}{(\textbf{Burnside's Lemma})}
    Let $G$ be a finite group acting on a finite set $X$. For each $g \in G$, define the fixed-point set
    \begin{equation}
        \textrm{Fix} (g) = \{ x \in X : g \cdot x = x  \}
    \end{equation}
    Then, the number of distinct orbits of $X$ under the action of $G$ is given by
    \begin{equation}
        \mid X/G \mid = \dfrac{1}{\mid G \mid} \sum_{g \in G} \mid \textrm{Fix} (g)  \mid.
    \end{equation}
\end{theorem}

\section{Proof of Theorem 3}
\label{Appendix:proof3}
\setcounter{theorem}{2}
\begin{theorem}
	The number of unique syndromes in the described QC-LDPC code is bounded by
	\begin{align}
		&\frac{1}{2^{m-\textrm{rank}(H)}} N_{\textrm{full}} \leq N_s \leq N_{\textrm{full}} \nonumber \\
		&N_{\textrm{full}} = \frac{1}{jp}\Bigl[\, 2^{jp} + (p-1)\,2^j + p\sum_{\substack{d\mid j\\d<j}}\varphi\!\bigl(\tfrac{j}{d}\bigr)\,2^{p\,d} \Bigr] \nonumber,
	\end{align}
	where $j$ and $p$ define the QC structure, $H$ is the parity-check matrix of the code, and $\varphi (\cdot)$ denotes the Euler's totient function.
	\label{th::unique_states}
\end{theorem}

\begin{proof}
    For each group element $g=\sigma^u \rho^s$, where $(0 \leq u < p,\, 0 \leq s < j)$, acting on the set $\Omega = \mathbb{Z}/p\mathbb{Z} \times  \mathbb{Z}/j\mathbb{Z} $, we analyze the resulting action to determine the number of cycles. The number of cycles depends on the specific values of $u$ and $s$. We divide the analysis into two cases, $s=0$ and $s \geq 1$. In the case where $s=0$, the group element is $g=\sigma^a$, which corresponds to a pure rotation by $u$ positions within each of the $j$ necklaces. If $u=0$, then $g$ is the identity transformation and fixes all $jp$ bead positions, resulting in $jp$ cycles. Hence, it contributes $2^{jp}$ to the sum. For $1 \leq u < p-1$, since $p$ is prime, we have $\gcd (u,p) = 1$, so $\sigma^u$ generates a single cycle of length $p$ on each necklace. Therefore, $\sigma^u$ exactly induces $j$ disjoint cycles on $\Omega$, contributing $2^j$ to the sum. As there are $p-1$ such nonzero values of $u$, the total contribution from all group elements of the form $\sigma^u$ is $2^{jp} + (p-1) 2^j$. In the case where $1 \leq s < j-1$, the group element is $g=\sigma^u \tau^s$. For any choice of $u$, the permutation $g$ breaks the full set of $jp$ beads into cycles of length $j/\gcd(s,j)$ in the necklace index and simultaneously that of length $p$ in the bead index. Hence, the total number of cycles is $p \gcd (s,j)$. Since $u$ ranges over $p$ values, the total contribution for a fixed $s$ is $p 2^{p \gcd (s, j)}$. Summing over all $s=1, 2, \cdots, j-1 $, the overall contribution from these elements is $p\sum_{s=1}^{j-1}2^{p\,\gcd(s,j)}$. From number theory, it is known that the sum over greatest common divisors can be rewritten using a divisor-sum identity $\sum_{s=1}^{j-1} 2^{p \gcd(s,j)} = \sum_{d|j, d<j} \varphi (j/d) 2^{pd}$, where $\varphi$ denotes the Euler's totient function. Finally, the number of distinct orbits under the group action is given by
    \begin{equation}
    \begin{split}
        |X/G|
        & = \frac{1}{jp}\Bigl[\,2^{jp} + (p-1)\,2^j + p\sum_{s=1}^{j-1}2^{p\,\gcd(s,j)}\Bigr] \\
        & = \frac{1}{jp}\Bigl[\, 2^{jp} + (p-1)\,2^j + p\sum_{\substack{d\mid j\\d<j}}\varphi\!\bigl(\tfrac{j}{d}\bigr)\,2^{p\,d} \Bigr].
    \end{split}
    \label{eq::burnside_Tanner}
    \end{equation}
    The number of distinct syndromes in the code space is $|S| = 2^{\textrm{rank} (H)} = 2^{n-k}$. Thus, the number of redundant rows in the parity-check matrix $H$ is $m - \textrm{rank} (H)$. 
    Thus, the counting is an overestimate, and $N_{\textrm{full}}$ provides an upper bound on the number of unique syndromes.
	Since the coloring of redundant rows can be identified through independent row colorings, the total number of distinct orbits for the group-structured QC-LDPC code can be lower bounded by $|X/G|/2^{m-\textrm{rank}(H)}$, where $|X/G|$ denotes the number of orbits under the group action, as given in Equation (\ref{eq::burnside_Tanner}).
    
\end{proof}


\end{document}